\documentclass[twocolumn,showpacs,preprintnumbers,amsmath,amssymb]{revtex4}
\usepackage{graphicx}

\begin{document}
\title{The flow and evolution of ice-sucrose crystal mushes}

\author{Andrew J. Gilbert}
 \email{ajg204@cam.ac.uk}
 \affiliation{Department of Earth Sciences, University of Cambridge}
\author{Felix K. Oppong}
 \email{felix.oppong@unilever.com}
 \affiliation{Unilever R\&D Colworth Science Park, MK44 1LQ}
\author{Robert S. Farr}
 \email{robert.farr@unilever.com}
 \affiliation{Unilever R\&D Colworth Science Park, MK44 1LQ}


\date{\today}

\begin{abstract}
We study the rheology of suspensions of ice crystals at moderate to
high volume fractions in a sucrose solution in which they are partially
soluble; a model system for a wide class of crystal mushes or slurries.
Under step changes in shear rate, the viscosity changes to a new
`relaxed' value over several minutes, in a manner well fitted by a 
single exponential. The behavior
of the relaxed viscosity is power-law shear thinning with shear rate, 
with an exponent of $-1.76 \pm 0.25$, so that shear stress falls
with increasing shear rate.
On longer timescales, the crystals ripen (leading to a falling viscosity)
so that the mean radius increases with time to the power $0.14 \pm 0.07$.
We speculate that this unusually small exponent is due to the
interaction of classical ripening dynamics with abrasion or breakup under flow.
We compare the rheological behavior to mechanistic
models based on flow-induced aggregation and breakup of crystal clusters,
finding that the exponents can be predicted from liquid
phase sintering and breakup by brittle fracture.
\end{abstract}

\maketitle

\section{Introduction}\label{sec:introduction}
The flow of crystal suspensions in which the crystals are soluble in the 
liquid phase (termed hereafter `mushes') is important in both nature and 
engineering. Examples from earth and planetary sciences involving this class of material include 
magmatic emplacement, lava flows \cite{Shaw,Caricci,Vona}, the formation of
sea ice \cite{Smedsrud} and cryogenic eruptions \cite{Lopes,Zhong,Ceres}. 
In an artificial setting they can occur in frozen foods \cite{Clark,Stokes},
slurry-ice refrigerant systems \cite{fridge}, 
metal casting \cite{casting1,casting2}, slurry explosives \cite{explosives}, solution 
mining \cite{mining} and evaporative mineral and sugar refinement.

Mushes consist of a suspension of hard, partially soluble crystals 
in a carrier liquid (which we term the `serum phase'); the latter
being usually Newtonian in rheology. 
However, additional complexity arises because the crystals have attractive 
or adhesive interactions resulting from van der Waals forces \cite{Hamaker}; but
more characteristically from their tendency to undergo liquid phase
sintering into clusters when they touch \cite{German}. The importance
of such attractive interactions will depend on the size and solubility of 
the crystals and the diffusivity of molecules in the serum, so one can
anticipate a spectrum of behaviors from approximately hard particle
suspension rheology to cases where sintering dominates.

As well as aggregation and sintering, soluble crystals will undergo ripening
\cite{Ostwald,LS,W}, driven by the minimization of interfacial energy as 
the larger crystals grow at the expense of the smaller. The mean crystal 
size will therefore gradually increase throughout any experiment.

These phenomena lead to complex rheological properties: mushes are
typically shear thinning at low solids volume fraction $\phi$, and can 
develop a yield stress and exhibit pseudo-plastic
behavior \cite{Stokes,Zhong,Kargel}. The time-dependent nature of cluster formation
and of sintering can also lead to strong history-dependence in the 
rheology. For example, if left unsheared, a mush will often
solidify \cite{Stokes} and in general one would expect it to display
thixotropic behavior \cite{thix} under changes in shear rate \cite{Zhong}.

The flow of a suspension of hard particles with no interactions 
other than hydrodynamics was 
first studied by Einstein \cite{Einstein}, who showed that spheres at a 
volume fraction $\phi \ll 1$ enhance the viscosity by a factor $(1+B\phi)$,
where $B$ is a pure number, termed the `intrinsic viscosity' or Einstein coefficient. 
For spheres this is 2.5, but is larger for other shapes \cite{Jeffery}. 
At higher volume fractions (particularly above $\phi=0.2$) the viscosity 
increases more strongly than the Einstein result \cite{Kruif} and will in 
general diverge at a maximum volume fraction $\phi_m$, which may be related 
to random close- \cite{Bernal} or loose- \cite{Song} packing, or result 
from a dynamic process of dilatancy \cite{RMP} or jamming \cite{FBM}. 
For the volume fractions of interest in this paper, a widely used 
approximation for the suspension viscosity $\eta$ in terms of the continuous 
phase viscosity $\eta_{0}$ is from Krieger and Dougherty \cite{KD}:
\begin{equation}
\eta = \eta_0 \left(1-\frac{\phi}{\phi_m}\right)^{-B\phi_m}, \label{eq:KrDo}
\end{equation}
where for subspherical particles of roughly equal size, one would 
use $B=2.5$ and $\phi_m \approx 0.64$.
Lower values of $\phi_m$ might be due to frictional interactions leading to
divergence at loose packing or lower packing density
due to significant departure from sphericity \cite{Delaney,Delaney2}.
Higher values may arise from polydispersity \cite{FarrGroot,FarrVisc}.

When attractive forces between particles are present, shear thinning and
other non-Newtonian behavior can result. For very dilute suspensions, 
aggregation leads to fractal flocs \cite{DLA,DLARLA}, which can ultimately 
percolate to form a gel. Before gelation, such suspensions are weakly shear 
thinning; while after, the gels are viscoelastic solids \cite{Buscall} with 
properties that scale with $\phi$.
At higher volume fractions (the subject of this paper), there is no
universal theory of attractive particle rheology and flow properties 
depend on the details of the interparticle interactions. 
Strongly cohesive suspensions can display steep shear thinning
and non-monotonic flow curves \cite{Buscall1,Buscall2}.

When making rheological measurements on such systems, 
pseudo-plasticity introduces problems in simple geometries. 
Using a parallel plate or Couette flow cell will lead to 
flow instabilities such as shear banding \cite{Schall}, while capillary
rheometers may be subject to plug flow \cite{Barnes}. These phenomena
leave the majority of the volume undeformed, so one is only probing the
flow in thin layers of the fluid. Moreover, with a history-dependent 
mush, the unsheared regions are likely to solidify, exacerbating the problem. 
In this paper, we therefore use a more complex geometry 
to ensure bulk deformation of the material. This consists of
a cylindrical vessel with a rotating impeller for which we measure
both the rotation rate $\omega$ and the applied torque $T$.

A second problem arises from the particle size of the suspension:
for reliable measurements, the flow geometry should ensure that the gaps
through which the suspension is forced to flow are large
compared to a crystal (or cluster), so that 
most of the viscous dissipation occurs in the bulk of the fluid
rather than in locally jammed regions in confined spaces \cite{Wyart}.
This, together with the size of the crystals and the values of the 
interesting shear rates, mean that we are not able to guarantee that the 
Reynolds number $Re$ of the flow is small. 

All these considerations mean that calibration of the rheological apparatus
is non-trivial, and for our case will rely on some specific assumptions 
about the rheology, which we are then able to test {\it post-hoc}. 

\section{Outline and key assumptions}

Throughout this paper, we suppose that under constant root-mean-square
shear rate $\dot{\gamma}_{\rm rms}$, the viscosity has a power law
dependence on this shear rate 
and the mean crystal radius $R$, with a volume fraction dependent pre-factor:
\begin{equation}\label{eq:eta_power_law}
\eta \propto \dot{\gamma}_{\rm rms}^{n_s}\, R^{n_r}.
\end{equation}
Here, $n_s$ and $n_r$ are exponents which we obtain in
section~\ref{sec:exponent_values}.
We hypothesize that this behavior arises from crystal clusters dynamically 
formed and broken up in the
flow. At high shear rates, we anticipate that the clusters will be broken
down, and there will be a crossover 
to Newtonian behavior with the Krieger-Dougherty value for the viscosity.

We also suppose that crystal ripening will occur and lead to a power-law 
dependence of crystal radius on time $t$, and potentially also on
root-mean-square shear rate 
(again with a volume fraction dependent prefactor):
\begin{equation}\label{eq:R_power_law}
R(\dot{\gamma}_{\rm rms},t) \propto t^{p_t}\, \dot{\gamma}_{\rm rms}^{p_s},
\end{equation}
where $p_t$ and $p_s$ are exponents we find in section~\ref{sec:size}.

Together, Eqs.\ (\ref{eq:eta_power_law}) and (\ref{eq:R_power_law})
imply that the observed viscosity in experiments at constant
volume fraction $\phi$, as a function of time and constant shear rate 
will have the following form:
\begin{equation}\label{eq:eta_power_law2}
\eta(\dot{\gamma}_{\rm rms},t) 
\propto t^{n_r \cdot p_t}\, \dot{\gamma}_{\rm rms}^{n_r \cdot p_s + n_s}.
\end{equation}
These exponent combinations are determined in sections~\ref{sec:continuous}
and~\ref{sec:size}.
The value for $n_s$ can be obtained more easily from experiments where
the shear rate is suddenly changed (see section~\ref{sec:sudden_change}).

The viscometer employed was first calibrated over a range of 
Reynolds numbers, using Newtonian fluids and hard sphere suspensions of
known viscosity, similar to the crystal mushes we are ultimately interested in.
The calibration (section~\ref{sec:calibration}) allows one to deduce viscosity from $T$ and
$\omega$ (and thus power dissipation), and also to attribute to
the flow root-mean-square values of shear rate $\dot{\gamma}_{\rm rms}$
and shear stress $\tau_{\rm rms}$.

In general, this calibration, performed for Newtonian fluids, 
cannot be used for non-Newtonian fluids.
This is because the material, being subject 
to different shear rates at different locations, will also have a 
spatially-dependent viscosity. The resulting flow pattern will not
correspond to any of the velocity fields covered by the Newtonian
calibration.

There is however a class of non-trivial rheologies to which the 
calibration does apply: suppose that following a sudden change in
shear rate, the viscosity is initially unchanged, but then relaxes
thixotropically towards a new viscosity corresponding to the new
shear rate. If this relaxation happens on a timescale that is long compared
to a rotation time of the viscometer, then the viscosity will remain
spatially uniform even as it changes slowly with time, and the Newtonian
calibration can be used to deduce the viscosity even as it relaxes 
to the new steady-state value.

The hallmark of this behavior is that if the rotation rate
is changed suddenly, the torque $T$ will
change discontinuously, but the calculated viscosity should be essentially 
continuous
(although its first derivative with time may be discontinuous). Furthermore,
only a relatively small fraction of the ultimate change
in viscosity should occur on a timescale of a single rotation of the impeller.
As we see in section~\ref{sec:sudden_change}, this is indeed 
observed for the crystal mushes studied here.

Microstructurally, we interpret this rheological behavior in the following 
way: if the dissipated power in the flow is coming primarily from the
deformation of the Newtonian serum, then high values of crystal
mush viscosity arise from the presence of crystal clusters
\cite{Caricci}, which are built up and break down dynamically in the flow. 
A sudden change in shear rate does not immediately 
affect the cluster statistics, so only acts
through a proportional change in the local flow rates in the serum
phase, and hence an `instantaneously Newtonian' behavior of the
suspension. Only over the course of several inverse shear rates does
the aggregation/breakup dynamics converge to a new distribution of
cluster sizes and shapes (and thus a new `relaxed viscosity').
Although both aggregation and breakup are likely to occur more quickly at 
higher shear rates, it is to be expected that breakup will increase more 
quickly than aggregation, so that cluster size and hence viscosity 
will fall with increasing shear rate. The relaxed
viscosity will therefore be shear thinning; a conclusion
borne out by our results in sections~\ref{sec:continuous} 
and~\ref{sec:sudden_change}.

A further complication is that even at constant rotation rate, the 
viscosity falls gradually over time, which we interpret to be due
to the slow growth by ripening of the crystals. Data on ripening
is obtained by optical microscopy on samples taken from the suspension,
and presented in section~\ref{sec:size}.

Finally, in section~\ref{sec:theory}, we make simple theoretical 
predictions for the relaxed viscosity
as a function of crystal size and suspension shear rate, based on
theories of adhesive contact or sintering and fracture of crystal
contacts, together with aggregation and breakup dynamics. 
To do this, we argue that at relatively high volume fractions,
and in the regime where viscosity is dominated by clustering,
the root mean square shear stress $\tau_{\rm rms}$ in the suspension is
\begin{equation}
\tau_{\rm rms} \propto \frac{F_{\max}}{R^2},
\end{equation}
where $F_{\max}$ is the force required to break an adhesive contact between
two crystals, and the dimensionless prefactor depends only weakly
on volume fraction.

This allows us to predict values for $n_s$ and
$n_r$ in terms of theories for adhesion, liquid phase sintering and
fracture, and compare these to the experimentally measured values.
The comparison of theory to experiment is shown in Fig.~\ref{fig:exponents},
and we find good agreement with a simple liquid phase
sintering model, and also the evaporation-condensation theory
of Kingery and Berg \cite{Kingery}.

\section{Materials and methods}

\begin{figure}
\begin{center}
\includegraphics[width=\columnwidth]{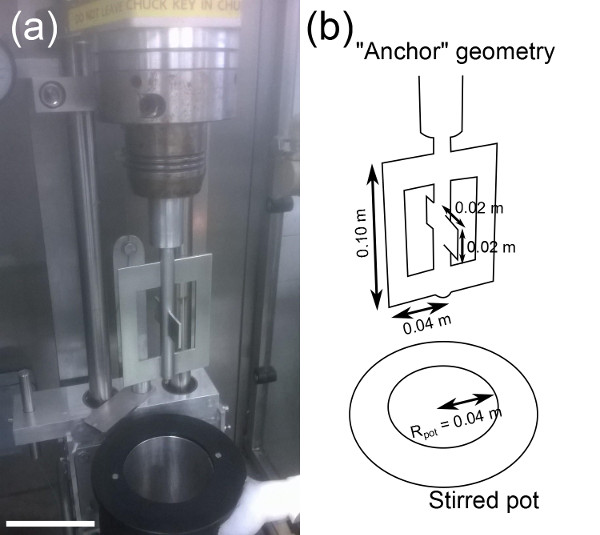}
\end{center}
\caption{(a) Image of `stirred pot' viscometer, with the `anchor' geometry impeller 
in the raised position. Scale bar is 10\,cm. (b) Line drawing of the anchor and stirred pot
showing dimensions of the anchor paddles and stirred pot. \label{fig:anchor}}
\end{figure}

The rheological apparatus used was a custom-made viscometer, developed by 
the engineering workshop at Unilever Research Colworth, and referred to
hereafter as a `stirred pot' (Fig. \ref{fig:anchor}). It consists of a
jacketed cylindrical vessel of inner radius $R_{\rm pot} = 0.04$\,m and depth 
$H_{\rm pot} = 0.135$\,m, maintained at a set temperature of 
$\Theta = -10^\circ$C by pumping propylene glycol solution through 
the jacket and a circulating, refrigerating Haake F8/C35 water bath. 
The impeller is an anchor geometry (see Fig.~\ref{fig:anchor}),  
which rotates at a chosen angular frequency $\omega$, while torque $T$ is recorded.
We chose rotation rates in the range 200 to 400\,rpm ($\omega =$ 20.9 to 41.9 rad s$^{-1}$).
In experiments with no change in rotation rate, data is collected 
every 300\,s over a run of several hours. In runs where we impose a step
change in rotation rate, data was collected every 2\,s, to observe the 
transients arising from this change.

All experiments on ice-sucrose mushes were conducted at fixed temperature 
$\Theta = -10^\circ$C $\pm 0.2^\circ$C, so the ice volume fraction 
$\phi$ is determined by the sucrose concentration only, 
and the unfrozen serum phase has a fixed composition
and therefore viscosity at this temperature. 
The sucrose concentrations used are shown in table~\ref{tab:formulations}, 
including a sucrose solution for which no crystals formed, hence for which $\phi = 0$, 
which is precisely the composition of the serum phase
in the ice-containing samples at this temperature. This zero-ice
formulation, being Newtonian, was used for one of the calibration
experiments. Sucrose solutions were prepared 
by adding boiled water to granulated sucrose, then
cooling overnight to $+5^\circ$C before pouring
into the stirred pot, which had been previously 
cooled to $-10^\circ$C.

The remaining calibration experiments were conducted at a range of different 
$\Theta$, and used 90\% and 99.5\% glycerol
($\rho = $1260\,kg\,m$^{-3}$, from Sigma-Aldrich); either alone, or as the serum phase
in a suspension of silica
spheres of $R = 100\,\mu$m (with a range of 50-150\,$\mu$m) and density 1800\,kg\,m$^{-3}$
(see Fig. \ref{fig:spheres}; also from Sigma-Aldrich). These suspensions
are predicted to be Newtonian and behave 
according to the Krieger-Dougherty relation, Eq.\ (\ref{eq:KrDo}).
The viscosities of the glycerol without silica spheres, and the sucrose 
solution without ice, were measured using an Anton Paar MCR501 rheometer.

In experiments where samples were extracted for image analysis, the stirrer 
was briefly stopped ($\approx 1$\,min), and a small portion of the sample was removed and the 
temperature checked. The torque was not affected by the stoppages, and returned to its previous value after being switched back on. The images were produced using a Leica 
DMLM microscope and Leica DFC490 camera, and the slides kept cool using a 
Linkam cold stage with the temperature regulated at $-10^\circ$C using a Landa RMB waterbath. 
The crystals were then analyzed using ImageJ software to get values of 
crystal radius and aspect ratio.

\begin{table}
\begin{center}
    \begin{tabular}{lll} 
    \hline\hline
    Sucrose (wt.\%)\hspace{1em} & Water (wt.\%)\hspace{1em} 
 & $\phi$ (ice) at -10$^{\circ}$C \\ \hline
    $55.0$ & $45.0$ & $0$ \\
    $41.7$ & $58.3$ & $0.33$ \\
    $37.1$ & $62.9$ & $0.42$ \\
    $32.5$ & $67.5$ & $0.51$ \\
    $28.8$ & $70.2$ & $0.57$ \\ 
    \hline\hline
    \end{tabular}
    \caption{\label{tab:formulations}Formulations of ice-sucrose suspensions, along with the ice volume fraction expected for each suspension at -$10^\circ$C.}
\end{center}
\end{table}

\section{Calibration} \label{sec:calibration}

The aim of our calibration is to be able to deduce viscosity $\eta$ from torque $T$
and rotation rate $\omega$. In order to do this we have performed a series of calibration
experiments using Newtonian suspensions where viscosity is known, 
$\omega$ is set and $T$ is measured. This allows us to come up with an
equation to predict suspension viscosity from the measured parameters of the system.

First, we define a Reynolds number for our system as: 
\begin{equation} \label{eq:Re}
Re \equiv \frac{\rho {R_{\rm pot}^2}\omega}{\eta},
\end{equation}
where $\rho$ is the suspension density.

We cannot assume the experiments are performed at low $Re$, with a laminar flow,
so the measurement equipment must be calibrated for a range of Reynolds numbers.
The key assumption, when we later use this calibration to determine mush 
viscosity, is that this viscosity is spatially uniform in the 
stirred pot, even if the material is thixotropic and (on longer timescales) 
shear thinning. The experimental justification for this assumption is 
provided in section~\ref{sec:sudden_change}.

\begin{figure}
\begin{center}
\includegraphics[width=\columnwidth]{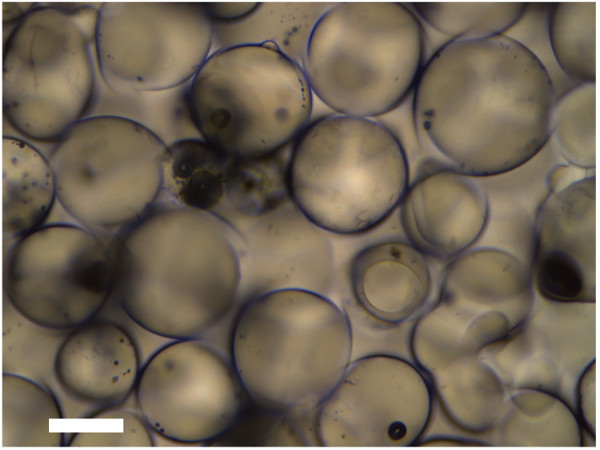}
\end{center}
\caption{Image of the silica spheres added to the glycerol. Scale bar 100$\mu$m. \label{fig:spheres}}
\end{figure}

The stirred pot was calibrated using 99.5\% glycerol at different
temperatures, either alone, or as the liquid phase in a suspension of
100$\,\mu$m radii silica spheres at +18$^\circ$C, +20$^\circ$C and 
+25$^\circ$C (see Fig.~\ref{fig:spheres}) and 90\% glycerol at +21$^\circ$C with no spheres. 
A second calibration set was performed using
55\% sucrose solution in water (with no silica spheres) at
$-10^\circ$C; a temperature and sucrose concentration where no ice is present 
(these systems are all Newtonian, or predicted to be so from Krieger-Dougherty). 
A third calibration set was performed using Lyle's Golden Syrup at +$25^\circ$C, this
set was performed to observe the limiting behaviour at low $Re$.
The viscosities of the glycerol and sucrose solutions were measured,
while viscosities of the silica sphere suspensions were deduced from these 
values and the Krieger-Dougherty relation for spheres, Eq.\ (\ref{eq:KrDo}),
using $B = 2.5$ and maximum packing fraction $\phi_m = 0.64$. 
The viscosity of golden syrup is 45\,Pa\,s at +25$^\circ$C \cite{Beckett}.

The glycerol and silica sphere suspensions were placed in the stirred pot,
and the torque $T$ and temperature $\Theta$ recorded as a function of
time $t$ at different angular velocities $\omega$. The silica sphere
suspension volume fractions used were 0, 0.42 and 0.51. 
For each run, measurements were taken over a time of 30\,mins, 
and were seen to be steady during that time.

The purpose of the calibration is to allow us to deduce the viscosity of
a fluid in the stirred pot from the torque and rotation rate.
Consider therefore the time-average power $P$ dissipated by viscous
flow in the stirred pot. Let $\langle\cdot\rangle$ denote an average 
over both space (within the pot) and time (over a few rotations), while
$\dot{\gamma}$ is the local, instantaneous shear rate in the suspension.
Then we note

\begin{equation} \label{eq:power_1}
P = T\omega = V_{\rm pot}\langle\eta \dot{\gamma}^2 \rangle
=V_{\rm pot} \eta {\dot{\gamma}_{\rm rms}}^2,
\end{equation}
where the volume occupied by the suspension is

\begin{equation}
V_{\rm pot} \approx\pi R_{\rm pot}^2 H_{\rm pot},
\end{equation}
and we have defined 
$\dot{\gamma}_{\rm rms}\equiv\langle\dot{\gamma}^2 \rangle^{1/2}$.
We have also used the assumption that the viscosity is uniform
and constant (over a few rotation times at least), to bring $\eta$ outside 
the spatio-temporal average.

\begin{figure}
\begin{center}
\includegraphics[width=\columnwidth]{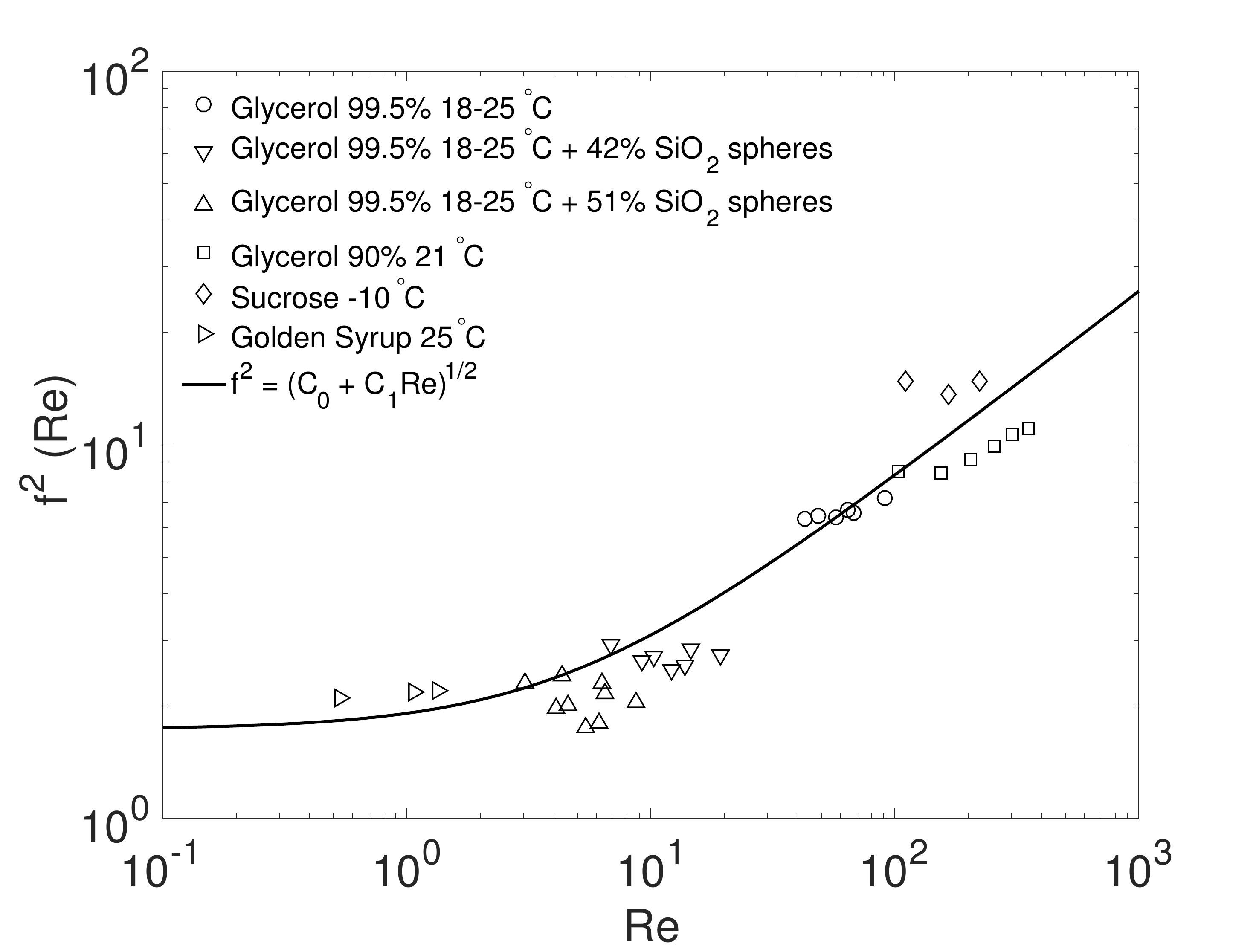}
\end{center}
\caption{Calibration curve showing how $f^2$ varies as a function of $Re$ for the various calibration experiments that have been performed. The curve is from Eq.\ (\ref{eq:f3}) using $C_0 = 3$ and $C_1 = 0.66$. \label{fig:calibration}}
\end{figure}

For small $Re \ll 1$ (creeping flow) we would expect 
$\dot{\gamma}_{\rm rms}\propto\omega$. However, for larger $Re \gg 1$, 
there may be a more complicated dependence. In general we take

\begin{equation}\label{eq:non_small}
\dot{\gamma}_{\rm rms} = f(Re)\omega ,
\end{equation}
for some function $f(Re)$ to be determined. From 
Eqs.\ (\ref{eq:power_1}) and (\ref{eq:non_small}) we find

\begin{equation}\label{eq:f2}
f^2 =\frac{T}{V_{\rm pot} \omega \eta},
\end{equation}
so that Eqs.\ (\ref{eq:Re}) and (\ref{eq:f2}) allow us to plot $f^2$ as a 
function of $Re$ for the calibration experiments,
as shown in Fig. \ref{fig:calibration}.

At low $Re$, $f^2$ must tend to a constant, while at higher $Re$, we find
$f^2 \propto Re^{\frac{1}{2}}$ (approximately).
We therefore fit $f^2$ to the following form:
\begin{equation} \label{eq:f3}
f^2 = (C_0 + C_1\,Re)^{\frac{1}{2}}.
\end{equation}
Plotting up the data in Fig.~\ref{fig:calibration}, we find that 
a good fit can be obtained with $C_0 = 3$ and $C_1 = 0.66$. 
Rearranging and solving the quadratic equation allows us to deduce a (spatially uniform) viscosity for the later 
experiments from known and measured quantities of $T$ and $\omega$:

\begin{equation} \label{eq:eta_calib}
\eta \approx \frac{T \chi}{0.33V_{\rm pot}\omega} \left[1+\left(1+\frac{3\chi^2}{0.33^2}\right)^{\frac{1}{2}}\right]^{-1}
\end{equation}
where $\chi$ is a non-dimensional quantity given as
\begin{equation} \label{eq:chi}
\chi \equiv \frac{T}{V_{\rm pot}\rho\omega^2 R_{\rm pot}^2}.
\end{equation}
We will also want to calculate 
values for $\dot{\gamma}_{\rm rms}$, which from Eq.\ (\ref{eq:power_1}) 
can be obtained from
\begin{equation}\label{eq:g_rms}
\dot{\gamma}_{\rm rms} = \left(
\frac{T\omega}{V_{\rm pot}\eta}\right)^{1/2}.
\end{equation}
Last, we define a root-mean-square stress
\begin{equation}\label{eq:tau_rms}
\tau_{\rm rms} \equiv \eta \dot{\gamma}_{\rm rms}.
\end{equation}

\section{Results for crystal mushes}

\subsection{Viscosity under constant rotation rate}\label{sec:continuous}

\begin{figure}
\begin{center}
\includegraphics[width=\columnwidth]{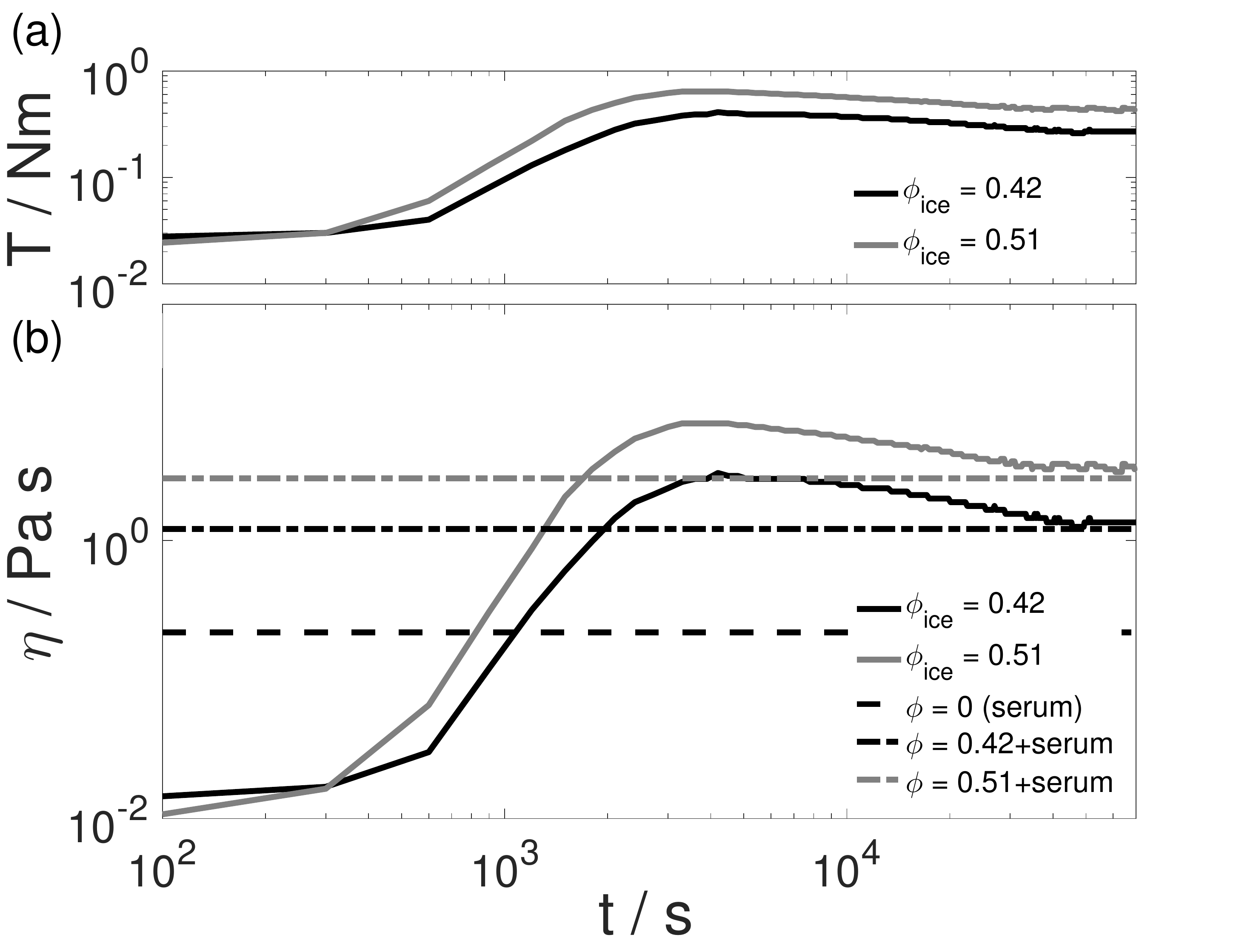}
\end{center}
\caption{(a) Variation in the torque profile as a function of time, 
for experiments where $\phi_{\rm ice} = 0.42$ and $\phi _{\rm ice}= 0.51$
at $\omega = 41.9$\,rad\,s$^{-1}$. (b) Profiles showing the change 
in viscosity over time as the solution 
crystallises. The black curve shows the change in viscosity 
for $\phi_{\rm ice} = 0.42$, while the gray curve shows the data 
for $\phi_{\rm ice} = 0.51$, both stirred at $\omega = 41.9$\,rad\,s$^{-1}$. 
The dash-dot lines show the expected viscosity for a sample 
with the same $\phi$ of hard spheres in equivalent serum. 
The dashed line shows the viscosity of the serum. 
The ice-sucrose suspension has a mean crystal size $R = 100\,\mu$m 
at $t = 13200$\,s. In these runs data was collected every $300$\,s.}
\label{fig:viscosity_time}
\end{figure}

Figures~\ref{fig:viscosity_time}(a) and~\ref{fig:viscosity_time}(b) 
show two typical profiles of torque and viscosity [deduced from Eq.\ 
(\ref{eq:eta_calib})] as a function of time for $\omega = 41.9$ rad s$^{-1}$ 
and $\phi_{\rm ice} = 0.42$ and $\phi_{\rm ice} = 0.51$.

In Fig.~\ref{fig:viscosity_time}(a) the measured torque starts low, 
as the stirred pot contains only a sucrose solution
at +5$^\circ$C, and no ice. The increase in torque (and viscosity) up to 
$t \approx$ 5000\,s represents both the cooling of
the solution and the build-up of ice volume fraction. After this,
both the temperature and ice content have reached steady-state values,
and it is the rheology from this time onwards that is the subject 
of the present paper.

Figure~\ref{fig:viscosity_time}(b) shows the calculated $\eta(t)$ for ice mushes with
$\phi_{\rm ice} = 0.42$ and $0.51$, as well as the viscosity of the sucrose solution 
serum phase at this temperature ($\phi=0$) and the predicted value of the
viscosity of a Krieger-Dougherty suspension in the same serum phase at 
volume fractions $\phi = 0.42$ and $0.51$. In all cases, the mush
viscosity is higher than the Krieger-Dougherty
predictions for hard, non-attractive spheres.

\begin{figure}
\begin{center}
\includegraphics[width=\columnwidth]{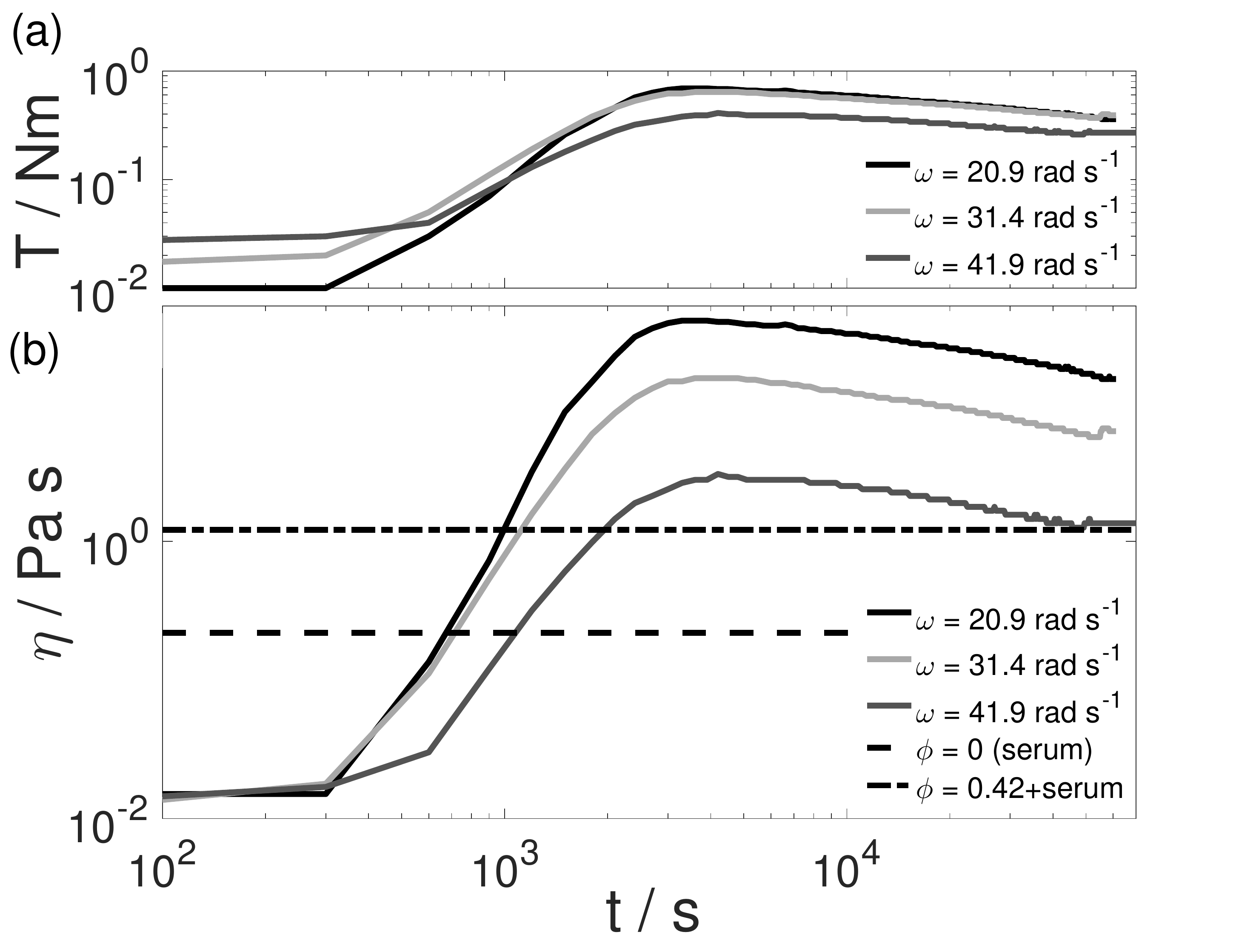}
\end{center}
\caption{(a) Profiles showing how torque varies through time for different values of $\omega$ for a suspension with $\phi_{\rm ice} = 0.42$. (b) Profiles showing how viscosity varies for the $\phi_{\rm ice} = 0.42$ samples with varied angular velocities. These experiments show the shear thinning nature of the ice-sucrose suspension. All shear rates show that the viscosity is greater than the suspension viscosity expected from Eq. (\ref{eq:KrDo}).}
\label{fig:viscosity_omega}
\end{figure}

The gradual decline of both torque and hence viscosity over the
remainder of the experiment is due to the increase of crystal size
through ripening. The decline of viscosity with time follows an approximate
power law, while there is strong decrease of viscosity with increasing
rotation rate $\omega$, shown in Fig.~\ref{fig:viscosity_omega}.
Between runs at different ice volume fractions we see an increase 
in viscosity with increasing $\phi$, however all runs reach peak 
$T$ (and hence $\eta$) at the same $t$. This is expected to be a result 
of clusters aggregating more efficiently under low shear conditions 
(due to the reduced force to break them up), and the increasingly
tortuous routes the serum must flow around the clusters. 
Shear thinning behavior is seen at higher $\omega$ as higher shear rates 
disrupt clusters, limiting the degree of aggregation that is possible 
and hence why high $\omega$ runs show viscosities closer to the 
Krieger-Dougherty prediction.

From these results, we observe that for the entirety of our run 
(post-crystallization peak) the viscosity is substantially larger 
than the Krieger-Dougherty result of non-interacting spheres 
(dash-dot lines on Figs.~\ref{fig:viscosity_time} and~\ref{fig:viscosity_omega}).
The viscosities decline with a power-law, suggesting that the 
suspension viscosity may eventually drop to that of the 
Krieger-Dougherty value and level off. With decreasing $\omega$ the 
time this levelling off takes is increased. 

\begin{figure}
\begin{center}
\includegraphics[width=0.6\columnwidth]{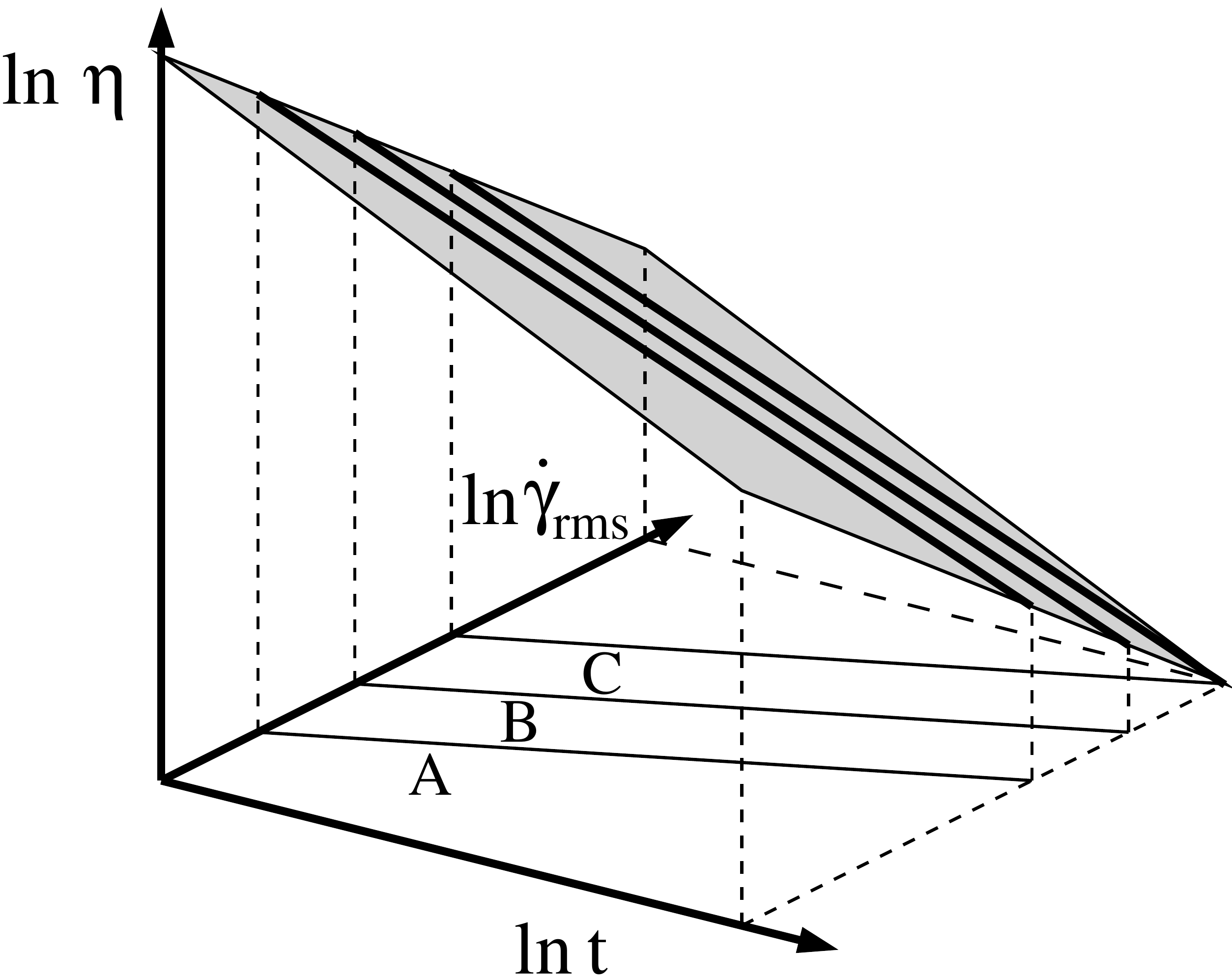}
\end{center}
\caption{Schematic of experimental procedure at constant rotation rate:
Several experiments (A, B and C in the figure) are performed
at constant rotation rate over a period of time.
This leads to changing $\dot{\gamma}_{\rm rms}$ over time $t$, and so
the combined set of data points for time, viscosity
$\eta$ and shear rate $\dot{\gamma}_{\rm rms}$ (the latter two
obtained from
calibration) fall on a plane (up to experimental error)
and are fitted to the power law of Eq.\ (\ref{eq:eta_power_law2})
using bivariate linear regression.
\label{fig:schematic_plane}}
\end{figure}

Ultimately, we are interested in the exponents $n_s$ and $n_r$ in 
Eq.\ (\ref{eq:eta_power_law}). However, in a given experiment 
at constant rotation rate, both $R$ and $\dot{\gamma}_{\rm rms}$
are changing (the latter because the viscosity changes as the crystals
grow in size). We therefore use Eq.\ (\ref{eq:eta_power_law2}) to obtain
the combinations of exponents $n_r\cdot p_t$ and
$n_r \cdot p_s + n_s$. Because both $t$ and
$\dot{\gamma}_{\rm rms}$ are changing, we need to perform 
a multivariate linear regression on runs at several different rotation
rates to obtain the requisite combinations of exponents.
The linear regression is shown schematically in 
Fig.~\ref{fig:schematic_plane}, and the results are shown in 
table~\ref{tab:measured_exponents_cts}.

Since in these experiments, when conducted at different shear rates,
the samples have undergone differing histories, we do not obtain
a simple measurement of the shear thinning exponent $n_s$ alone. This
can be seen in Eq.\ (\ref{eq:eta_power_law2}), where it is
only the combination of exponents that is accessible. To remedy this,
we look in the next subsection at experiments where a sample is
sheared at a constant initial rate $\omega_i$, and then at a consistent
time in the experiment, the shear rate is stepped to a new value $\omega_f$.

\begin{table*}
\begin{center}
    \begin{tabular}{lllccc} 
    \hline\hline
    $\phi$ (ice) & values of $\omega$\,/\,rad\,s$^{-1}$
       & $p_t$ & $p_s$ & $(n_r \cdot p_t)$ & $(n_r\cdot p_s + n_s)$ \\ \hline
    $0.42$ & 20.9, 31.4, 41.9 & $0.14 \pm 0.07$ & $-0.01 \pm 0.19$ & $-0.25\pm0.20$ & $-1.72 \pm 0.20$. \\
    $0.51$ & 20.9, 31.4, 41.9 & n.m. & n.m. & $-0.24\pm0.20$ & $-1.25\pm0.20$. \\
    $0.57$ & 20.9, 31.4 & n.m. & n.m. & $-0.19\pm0.20$ & $-1.51\pm 0.20$. \\
    \hline\hline
    \end{tabular}
    \caption{\label{tab:measured_exponents_cts}Experimental values for 
combinations of exponents [see Eqs.\ (\ref{eq:R_power_law})
and (\ref{eq:eta_power_law2}) in the text]
for mushes under constant rotation rate $\omega$ 
(but varying $\dot{\gamma}_{\rm rms}$).
The results are obtained from multivariate linear regression, fitting either
$\ln\eta$ or $\ln R$ to a linear function of $\ln t$ 
and $\ln\dot{\gamma}_{\rm rms}$. Several experiments, at differring
but constant $\omega$ are used for each regression analysis.
Some quantities are not measured in particular 
experiments; these are denoted `n.m.'.}
\end{center}
\end{table*}

\subsection{Viscosity after a step change in rotation rate}\label{sec:sudden_change}

\begin{figure}
\begin{center}
\includegraphics[width=\columnwidth]{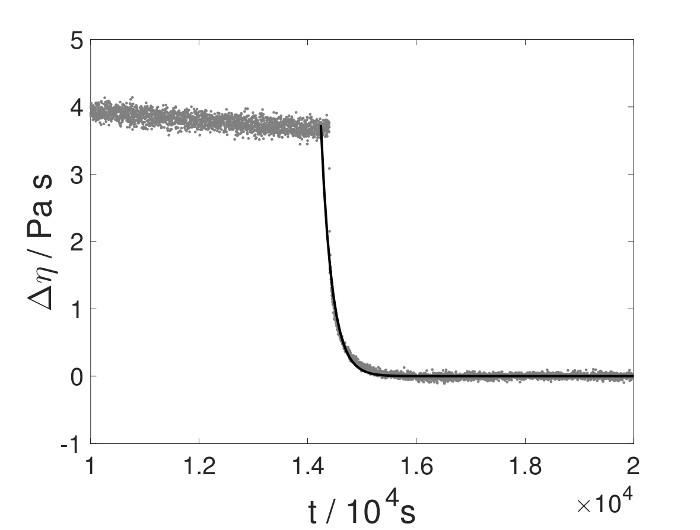}
\end{center}
\caption{Exponential decay of $\Delta\eta$ (gray points) as a function 
of time before and after a step change in rotation rate, where
$\Delta\eta$ is the difference between the calculated viscosity
and the relaxed viscosity at the new shear rate, extrapolated back to the
time the rotation rate was changed.
For this experiment $\phi_{\rm ice} = 0.42$, $\omega_i = 31.4$\,rad\,s$^{-1}$, 
and $\omega_f = 41.9$\,rad\,s$^{-1}$, with the step change in
$\omega$ occurring at $t = 14400$\,s. 
The black dashed curve is the best fit exponential
with the form $\Delta\eta = \Delta\eta_0~\exp(-t/t_{\rm relax})$, 
and from this $t_{\rm relax}$ can be determined.
\label{fig:exponential}}
\end{figure}

In section~\ref{sec:calibration}, for our calibration to be correct, 
our mushes are expected to see an instantaneously 
Newtonian response of the suspension to changes in shear rate. 
This is followed by a thixotropic relaxation period 
as the clusters re-organize and adapt to their new shear environment.
A series of experiments have been run to observe this behavior.
Samples are prepared in the same way as for the continuous experiments, 
and placed in the stirred pot, set at $\omega_i = 31.4$\,rad\,s$^{-1}$ 
and $\Theta = -10^\circ$C. Then, at $t = $14400\,s $\omega$ is rapidly 
raised or lowered to a new rotation rate $\omega_f$.
The suspension is then allowed to relax with no further changes in $\omega$.
The impeller takes around 10\,s to get up to the correct speed when the
set-point is changed and, as Fig.~\ref{fig:exponential} shows, 
the relaxation timescale is considerably greater than 10\,s.

\begin{figure*} 
\begin{center}
\includegraphics[width=\textwidth]{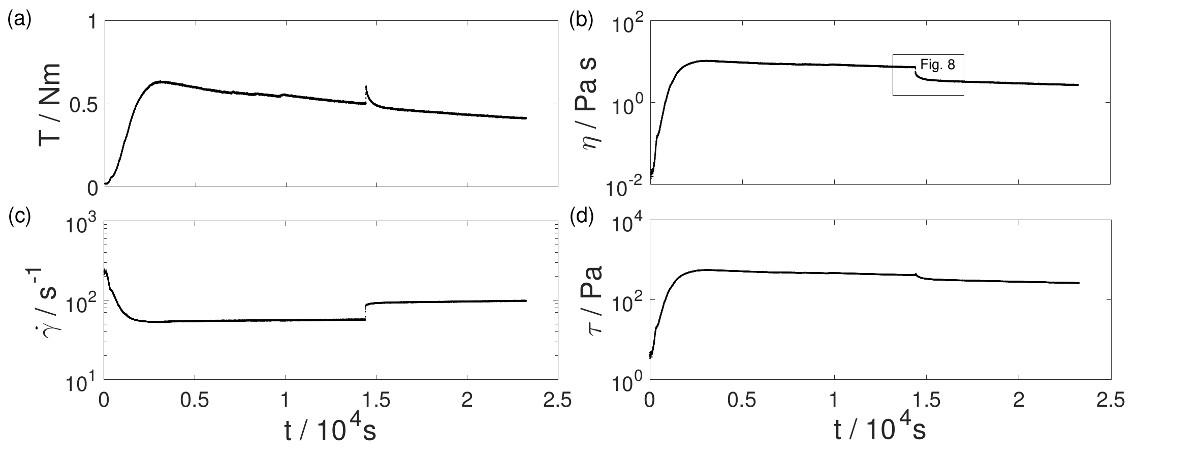}
\end{center}
\caption{Measured and calculated properties and how they evolve 
through a step change in $\omega$.
For this experiment $\phi_{\rm ice} = 0.42$, $\omega_i = 31.4$\,rad\,s$^{-1}$, 
and $\omega_f = 41.9$\,rad\,s$^{-1}$, with the step change at $t = 14400$\,s. 
(a) The measured torque changes discontinuously, with a sudden sharp increase when $\omega$ is dropped, before relaxing to below the original value. (b) $\eta$ [calculated from Eq.\ (\ref{eq:eta_calib})] changes more continuously and shows a thixotropic relaxation following the change in $\omega$. Rectangle denotes portion which is shown in detail in fig.~\ref{fig:exponential}. (c) $\dot{\gamma}_{\rm rms}$ shows a jump before steadying. This value is calculated from Eq.\ (\ref{eq:g_rms}). (d) Overall root-mean-square stress shows a small decrease due to the change in $\omega$, calculated from Eq.\ (\ref{eq:tau_rms}). In this experiment data is logged every 2\,s.
\label{fig:step_expt}}
\end{figure*}

\begin{figure}
\begin{center}
\includegraphics[width=\columnwidth]{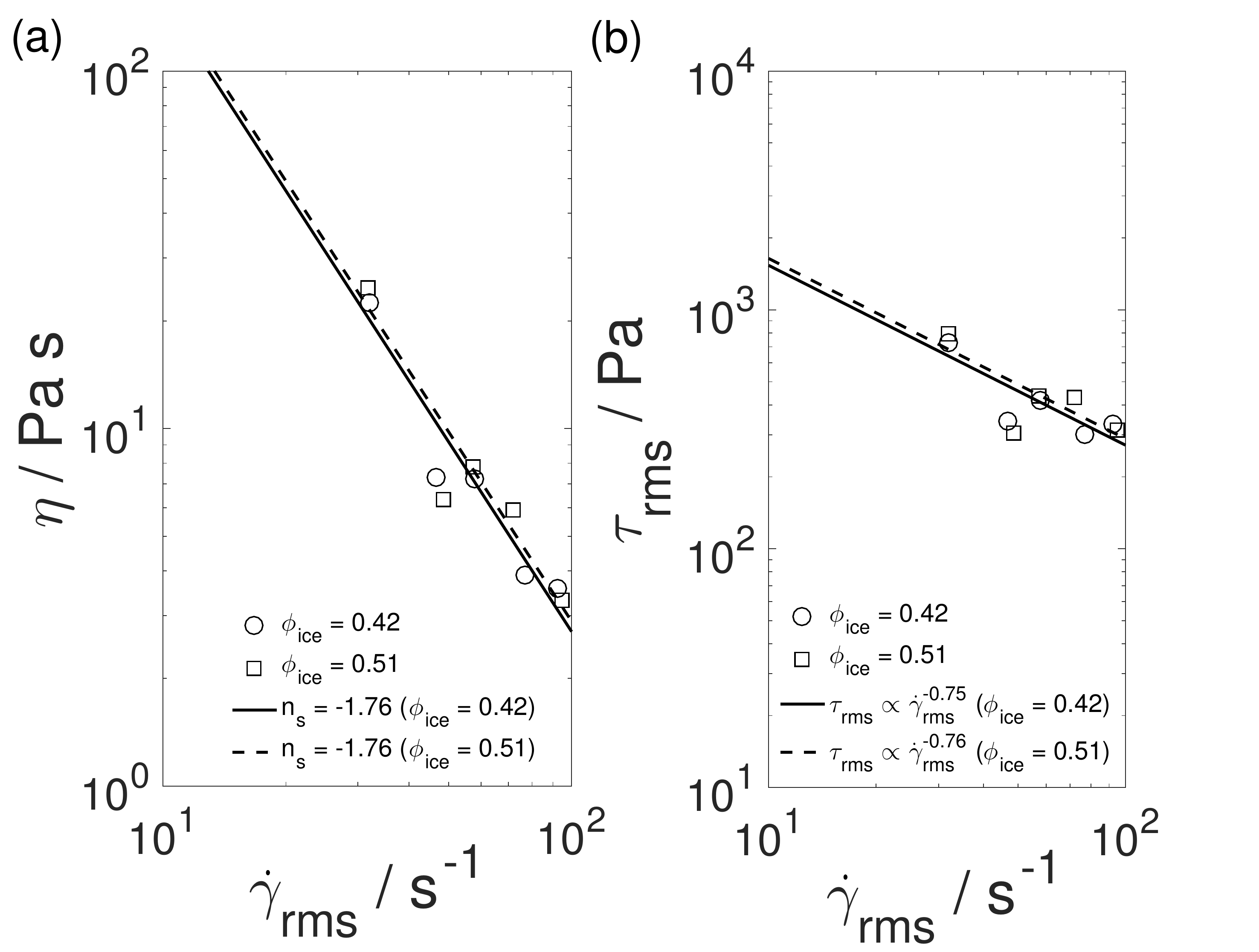}
\end{center}
\caption{(a) Plot showing the measured viscosity and root-mean-square shear 
rate at $t = 14400$\,s. The lines are best fit lines, with prefactors related 
to $\phi$ and the exponents given 
as $n_s = -1.76 \pm 0.23$ for $\phi_{\rm ice} = 0.42$ 
and $n_s = 1.76 \pm 0.27$ for $\phi_{\rm ice} = 0.51$. 
(b) Plot showing the root-mean-square shear stress versus root-mean-square 
shear rate at $t = 14400$\,s. Both lines have a gradient of $-0.75$.
For both plots error bars are the size of the markers. 
\label{fig:eta_gamma}}
\end{figure}

The results of the experiments show that although $T$ changes discontinuously, 
$\eta$ changes continuously (albeit inertia of the experimental 
apparatus is a potentially confounding effect). The viscosity
displays a shear-thinning 
and thixotropic response [see Figs.~\ref{fig:step_expt}(a)-(d) and~\ref{fig:exponential} for 
an increase in $\omega$], with a relaxation timescale $t_{\rm relax} 
\approx 300$\,s much greater than the time for a single rotation of the impeller. 
From analyzing the instantaneous
$T$, $\eta$ and $\dot{\gamma}_{\rm rms}$ at $t = 14400$\,s, and extrapolating to that time 
for the relaxed response, we can calculate $\eta$, $\dot{\gamma}_{\rm rms}$ and
$\tau_{\rm rms}$ at the step-point for both values of $\omega$.
Plotting these data points in Figs.~\ref{fig:eta_gamma}(a)
and~\ref{fig:eta_gamma}(b) we evaluate $n_s$ from Eq. (\ref{eq:eta_power_law}) 
at $t =$ 14400\,s. Since $R$ is constant, $n_r$ can be disregarded in this 
calculation and we find an average value of $n_s = -1.76 \pm 0.25$. 

We can calculate the timescale of thixotropic relaxation $t_{\rm relax}$ 
by fitting an exponential curve to the relaxation period 
where the long-term trend has been removed. 
This is shown in Fig.~\ref{fig:exponential} and average values of 
$t_{\rm relax} =$ 370\,s for both values of $\phi_{\rm ice}$ are found in table~\ref{tab:measured_exponents_step}.

\begin{table}
\begin{center}
    \begin{tabular}{lcccc} 
    \hline\hline
    $\phi_{\rm ice}$ \hspace{1em}
       & $\omega_i$\,/\,rad\,s$^{-1}$
       & $\omega_f$\,/\,rad\,s$^{-1}$
       & $t_{\rm relax}$/s\\ \hline
     $0.42$ & $31.4$ & $20.9$ & $204\pm6$ \\
     $0.42$ & $31.4$ & $26.2$ & $354\pm5$ \\
     $0.42$ & $31.4$ & $36.7$ & $554\pm9$ \\   
     $0.42$ & $31.4$ & $41.9$ & $622\pm13$ \\
     $0.51$ & $31.4$ & $20.9$ & $400\pm10$ \\
     $0.51$ & $31.4$ & $26.2$ & $391\pm7$ \\
     $0.51$ & $31.4$ & $36.7$ & $182\pm8$ \\   
     $0.51$ & $31.4$ & $41.9$ & $243\pm13$ \\
    \hline\hline
    \end{tabular}
    \caption{\label{tab:measured_exponents_step}Experimental values for the
thixotropic viscosity relaxation time $t_{\rm relax}$ calculated from
the fitted exponential decay. Results are from 
experiments where 
the rotation rate starts at $\omega_i$, and at a time $t = 14400$\,s is suddenly 
changed to a new value $\omega_f$.}
\end{center}
\end{table}

\subsection{Crystal radius and shape}\label{sec:size}
\begin{figure}
\begin{center}
\includegraphics[width=\columnwidth]{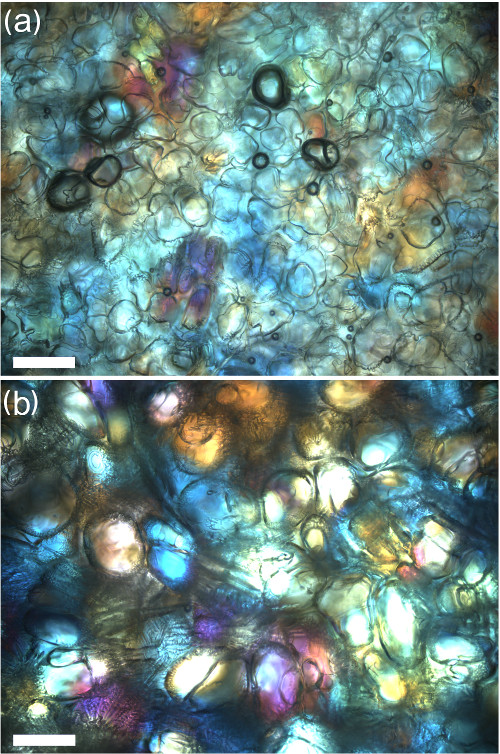}
\end{center}
\caption{Images of ice crystals after different residence times within the 
stirred pot, in a $\phi_{\rm ice} = 0.42$ suspension, with $\omega = 31.4$\,rad\,s$^{-1}$. 
(a) After $t =$ 4230\,s (corresponding to the time of peak 
torque). (b) After $t =$ 83000s\,. Scale bar for each is 200$\,\mu$m.
\label{ice_crystals}}
\end{figure}

The size and shape of the ice crystals produced were measured at various 
times during each run. The average radius $R$ of the ice crystals increases 
with time $t$, as seen in Fig~\ref{ice_crystals}(a) 
and ~\ref{ice_crystals}(b). 

At a fixed $T$, one might expect $R$
to increase as a power of time $t$, and also to depend on ice volume fraction
and perhaps shear rate, as shown in Eq.\ (\ref{eq:R_power_law}).
We believe that the secular increase in crystal size throughout an 
experiment at constant
$\omega$ accounts for the slow decline in torque and (calculated) viscosity.

At zero shear rate, if crystals grow by Ostwald 
ripening \cite{Ostwald}, driven by the Laplace pressure of the 
ice/liquid interfacial energy $\sigma$ and limited by the interdiffusivity 
of solute and solvent through the liquid, then one would expect the
ripening exponent $p_t = 1/3$.
This scaling has been observed not only in the dilute limit 
where Lifshitz-Slyozov-Wagner (LSW \cite{LS,W}) theory applies, 
but also in more concentrated systems \cite{Bremer}. 
One would expect shear to accelerate the ripening by introducing mixing 
in addition to diffusion, but the only mechanism which would lead to a change
of exponent $p_t$, is that if interfacial attachment kinetics can
limit ripening, rather than diffusion, then a value of $p_t = 1/2$ would
result \cite{W}. Mixing from shear might 
push a system into this regime from the diffusion-limited case.

\begin{figure}
\begin{center}
\includegraphics[width=\columnwidth]{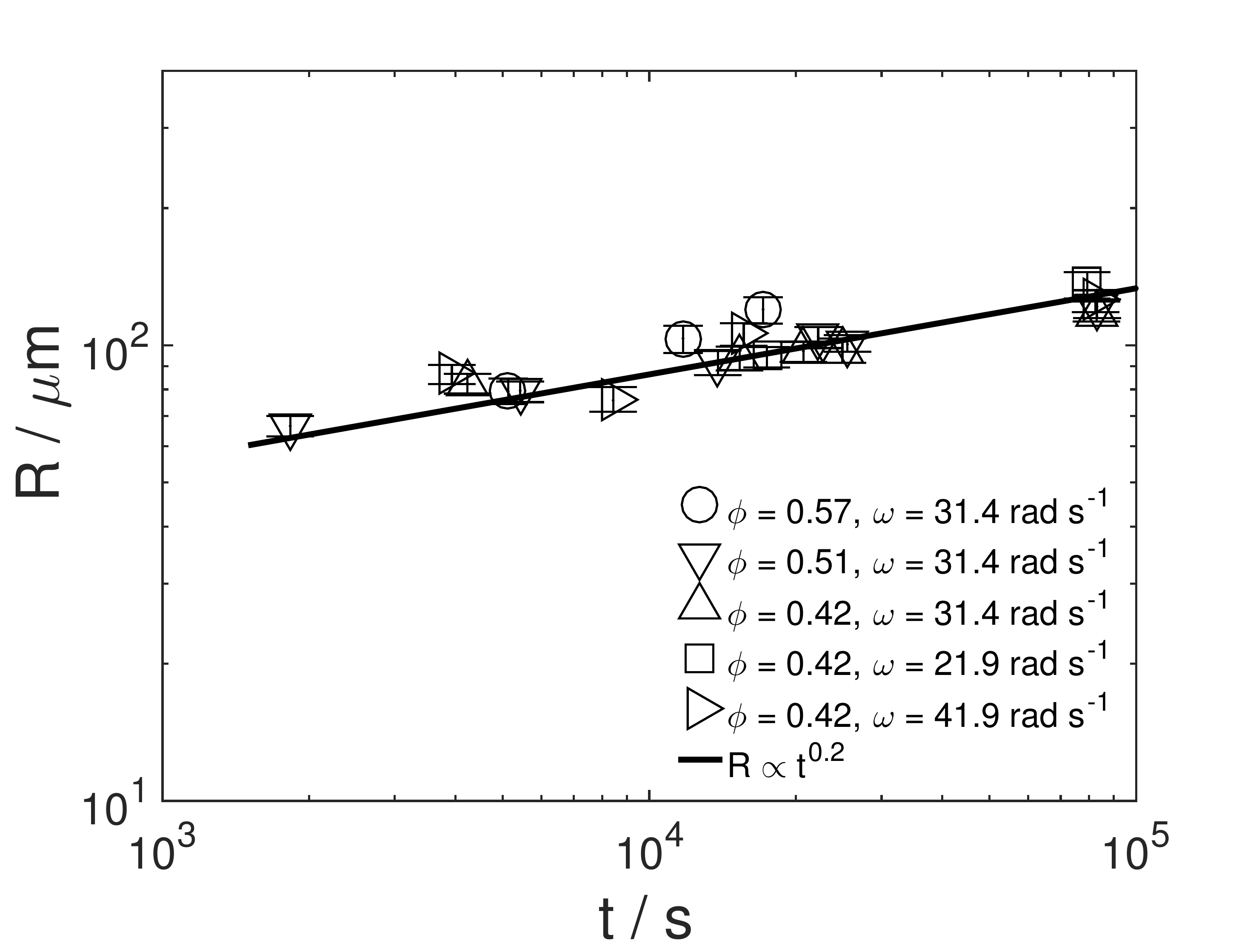}
\end{center}
\caption{Changes in crystal radius as a function of time in the stirred pot. 
Solid line is best fit power law, with slope $0.2$. 
Error bars show 95\% confidence intervals, calculated using a bootstrap method.
\label{fig:radius_time}}
\end{figure}

However, from imaging the crystals in the mushes studied here, we observe 
for a range of conditions (see fig.~\ref{fig:radius_time})
that $R \propto t^{0.2}$ at fixed $\omega$, and depends only weakly
on rotation rate.
In order to obtain the exponents $p_t$ and $p_s$ in 
Eq.\ (\ref{eq:R_power_law}), we again note that at constant
$\omega$, the shear rate will vary with time, so we need to
perform bivariate regression analysis, using a range of steady values 
of $\omega$, and fitting $\ln R$ to a linear combination of
$\ln t$ and $\ln\dot{\gamma}_{\rm rms}$.

The results, shown in table \ref{tab:measured_exponents_cts},
are that $p_s$ is small (indeed consistent with being zero); while
$p_t$ takes the unexpectedly low value $p_t = 0.14 \pm 0.07$. 
(Figure~\ref{fig:radius_time} shows a best-fit line with exponent $= 0.2$ 
for all crystal sizes; but this includes some data that was excluded from the
bivariate regression analysis since we only have 
different $\omega$ experiments for $\phi = 0.42$).
A low value for the exponent $p_t$ (relative to theories
of ripening) has also been seen previously
in cryogenic ripening under shear \cite{Pronk}, where ice has been 
observed crystallizing within a NaCl solution at different temperatures 
and shear rates. However, to our knowledge, no theory has predicted 
this exponent for ripening crystal systems.

\begin{figure}
\begin{center}
\includegraphics[width=\columnwidth]{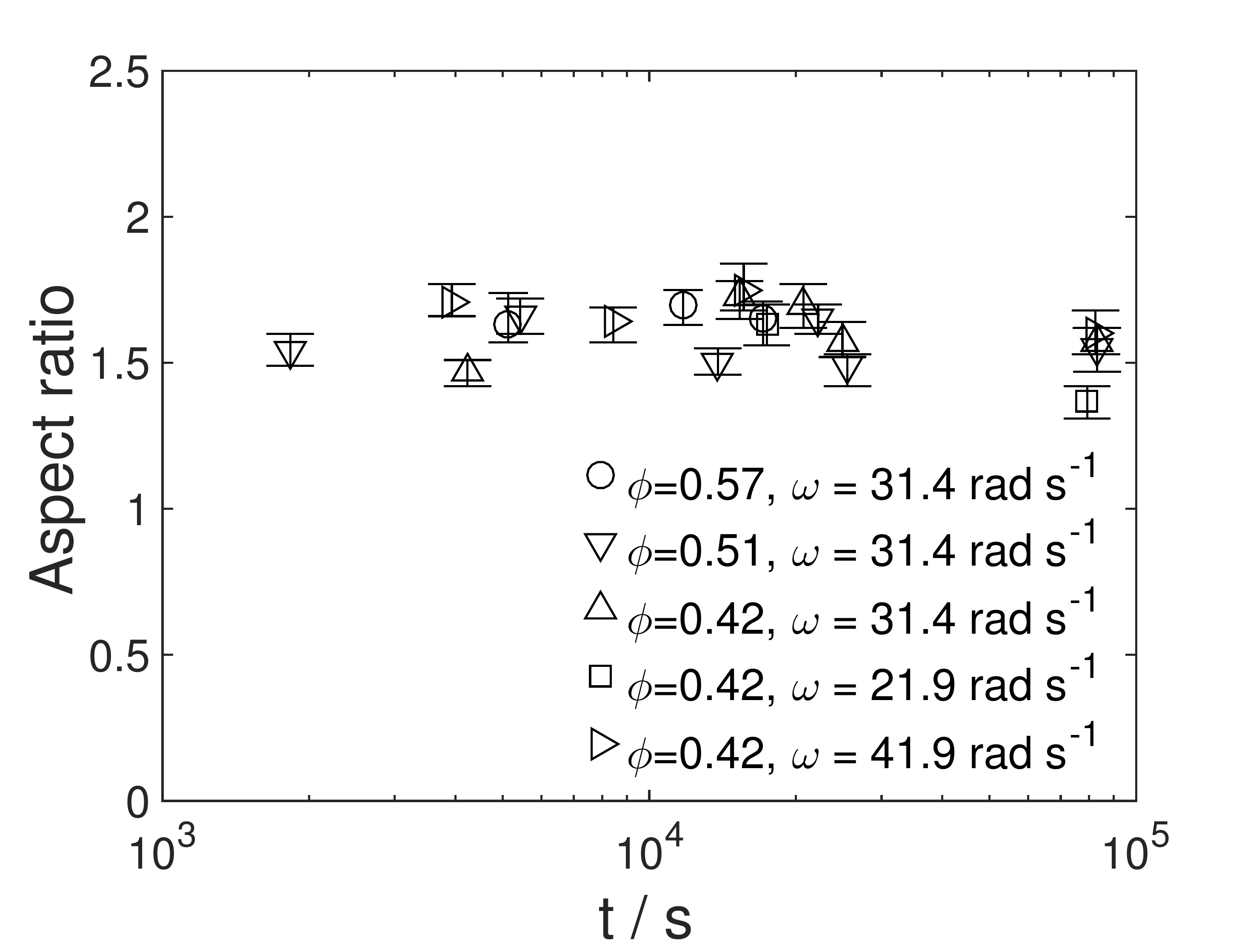}
\end{center}
\caption{Aspect ratio of ice crystals as a function of time in the stirred pot. Error bars show 95\% confidence intervals, calculated using a bootstrap method.
\label{fig:aspect_time}}
\end{figure}

As well as size, we measured the aspect ratios of crystals sampled from the
stirred pot. The shape of the crystals does not change significantly 
with time, with the aspect ratio remaining around 1.6, as shown in 
Fig.~\ref{fig:aspect_time} (the same result as observed by Ref.\ \cite{Pronk}). 

\begin{figure}
\begin{center}
\includegraphics[width=\columnwidth]{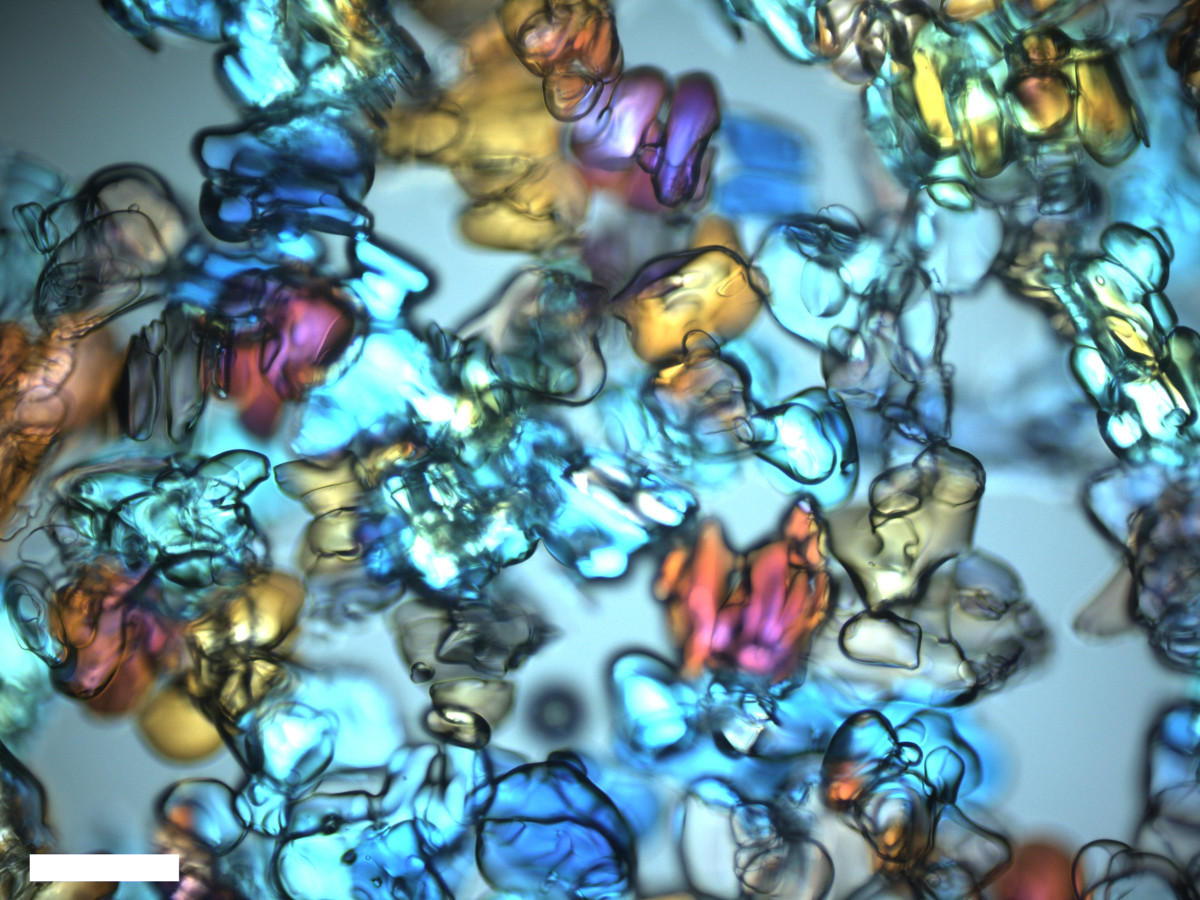}
\end{center}
\caption{Image of crystals in $\phi_{\rm ice} = 0.42$ at $\omega = 31.4$\,rad\,s$^{-1}$, at $t = 25000$\,s. There are numerous misshapen grains, with some showing instances of abrasion and damage. Scale bar is 200\,${\mu}$m.
\label{fig:abrasion}}
\end{figure}

In some images of ice however, the crystals appeared to be abraded and had 
rough surfaces and were often misshapen (this can be seen in Fig.~\ref{fig:abrasion}). 
Abrasion and breaking of partially sintered 
grains might be expected to occur under a high shear rate, so these processes 
are a possible cause for the unexpectedly small growth 
exponent $p_t$ we observed. 

\subsection{Values and consistency of the exponents}\label{sec:exponent_values}

From the results in tables~\ref{tab:measured_exponents_cts} and
Fig.~\ref{fig:eta_gamma}(a) we deduce the four exponents, 
that are predicted in Eqs.\ (\ref{eq:eta_power_law})--(\ref{eq:eta_power_law2})
using the step change experiments for $n_s$ and the constant $\omega$
experiments for $p_t$, $p_s$ and $n_r \cdot p_t$ to obtain:
\begin{eqnarray}
n_s &=& -1.76 \pm 0.25 , \label{eq:ns} \\
n_r &=& -1.8 \pm 1.3 , \label{eq:nr} \\
p_t &=& 0.14 \pm 0.07 , \label{eq:pt} \\
p_s &=& -0.01 \pm 0.19 \label{eq:ps} .
\end{eqnarray}
We also have an independent test of these exponents because
we measure $n_r\cdot p_s + n_s$, but we have not used this in deriving
Eqs.\ (\ref{eq:ns})--(\ref{eq:ps}). We find, from these equations,
$n_r\cdot p_s + n_s = -1.78 \pm 0.34$. This is consistent with the measured
value in table~\ref{tab:measured_exponents_cts} of $-1.49 \pm 0.12$.

\section{Theoretical models}\label{sec:theory}

\subsection{Stress arising from dynamic clusters}

In this section, we relate our observed rheology to the microstructural
physics of the mushes.

At moderate to high volume fraction, the presence of even simple hard
particles substantially increases the viscosity of a suspension over
that of the serum. For our systems, the introduction of adhesive forces
leads to a further large increase in viscosity, as seen in
Fig.~\ref{fig:viscosity_omega}.
There are two complementary ways to view this:
through power dissipation and through force networks.

In the view based on power dissipation, for hard particles,
this power is generated exclusively in the solvent.
This remains essentially true if there are adhesive contacts,
as long as they are brittle, so they break at very
small strains. The increased viscosity arises from the tortuous paths
and higher local shear rates imposed upon the serum as it flows
around the particles. Even in the absence of adhesive forces,
the flow must generate some correlations in particle position \cite{Marrucci},
otherwise a simple self-consistent picture of particle pairs
passing one another would predict a logarithmic divergence of
viscosity with $(\phi-\phi_m)$, in contrast
to the much stronger divergence predicted in Eq.\ (\ref{eq:KrDo}).
The high viscosities associated with adhesive forces must arise, in this
picture, from the formation of extended structures (clusters) of crystals
in the flow, which force even larger local strain rates on the serum phase.

The second picture is based on force networks. At high volume fractions,
hydrodynamic forces become more localized between neighboring particles,
so to a good approximation one can ignore the serum replacing it with 
pairwise lubrication forces between the crystals.

The key assumption we make in this section is that the clusters
formed in the flow are transient and in a quasi-steady state
(ignoring the slow ripening dynamics). That is to say they form and
break up over a timescale of order an inverse shear rate. More
specifically, we suppose that any bond which forms between a
pair of crystals survives only for a time 
$t_{\rm bond}\sim \dot{\gamma}_{\rm rms}^{-1}$ before being ruptured.
This means that if there are clusters present,
a typical tensional force between crystal pairs, in the extensional
direction of the flow, is of order the rupture force $F_{\max}$
of the adhesive bond that is present. We assume that this sets
the scale for all interparticle forces (extensional or compressive)
in the flow.

Let the $i$'th component of the pair force between particles $m$ 
and $n$ be $F^{[m,n]}_{i}$, where the $j$'th component of the vector 
joining their centers is $r^{[m,n]}_{j}$, and let the representative volume of the suspension 
under consideration be $V$. The mean stress tensor in the suspension, with 
components $\tau_{i,j}$ will be given by \cite{Allen}

\begin{equation}\label{eq:tensor}
\tau_{i,j}=\frac{1}{2V}\sum_{m,n}F^{[m,n]}_{i}r^{[m,n]}_{j}.
\end{equation}

Since contact forces are only possible between near neighbors, 
the number of pairs of particles per unit volume between which a 
non-zero force obtains will scale as $R^{-3}$ (with a prefactor
that depends slightly on volume fraction near $\phi_m$).
For moderate-to-high volume fractions, the length of each 
vector ${\rm\bf r}^{[m,n]}$ will be close to $2R$, and
by assumption the forces are of typical magnitude $F_{\max}$,
the rupture force of an adhesive bond.
So from Eq.\ (\ref{eq:tensor}), we arrive at an estimate for the
root-mean-square shear stress in the system:
\begin{equation}\label{eq:tau_estimate}
\tau_{\rm rms} \propto F_{\max} R^{-2},
\end{equation}
where the constant of proportionality is dimensionless, of order unity,
and depends only weakly on volume fraction near $\phi_m$.

We expect Eq.\ (\ref{eq:tau_estimate}) to hold while there is
a substantial amount of (transient) adhesive clusters in the flow.
At high enough shear rates, it is possible that breakup is
so effective that all adhesive bonds are broken in a time much
less than $\dot{\gamma}_{\rm rms}^{-1}$, so there are effectively
no clusters. As this state is approached, there will be a crossover to
the hard particle (Krieger-Dougherty) viscosity.

The problem now reduces to finding the maximum force required to separate 
two crystals after they are brought into contact. We expect this to 
depend in a power-law manner on the crystal radius $R$ and the
contact time $t_{\rm bond}\sim\dot{\gamma}_{\rm rms}^{-1}$. 
Substituting such a power law
behavior into Eq.\ (\ref{eq:tau_estimate}), and comparing to 
Eq.\ (\ref{eq:eta_power_law}) allows us to relate the exponents 
(which will be predicted by the various theories that follow) to
$n_s$ and $n_r$:

\begin{equation}\label{eq:Fmax}
F_{\max} \propto \dot{\gamma}_{\rm rms}^{n_s + 1} R^{n_r + 2}.
\end{equation}

In the following subsection, we present some theories from the literature 
for the bond strength $F_{\rm max}$ in terms of $t_{\rm bond}$ and $R$.

\subsection{Theory for adhesive spheres}

Johnson-Kendall-Roberts (JKR) theory \cite{JKR} describes the contact 
mechanics between two elastic spheres of radius $R$ when there is a 
reversible, adhesive interfacial energy per unit area $\sigma_{\rm surf}$ 
between them if they touch. The resulting expression for the maximum force 
to separate them is
\begin{equation}
F_{\max} = \frac{3\pi}{2}\sigma_{\rm surf}R.
\end{equation}
From Eq.\ (\ref{eq:Fmax}), this leads to predicted exponents $n_s = n_r = -1$.

Direct measurements of adhesive force have been performed on 
micro-manipulated ice particles in air and sucrose solution \cite{Fan}.
These results are broadly consistent with JKR theory, but display a
time dependence not present in the theory.

\begin{figure}
\begin{center}
\includegraphics[width=\columnwidth]{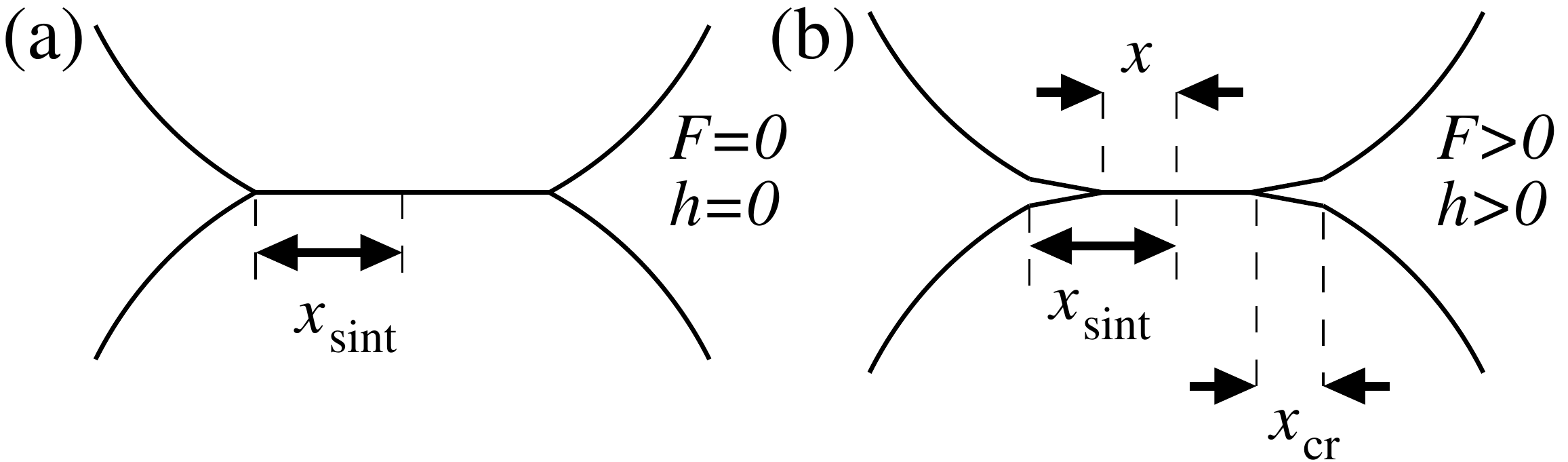}
\end{center}
\caption{(a) Geometry of two spheres of radius $R$ which have come into 
contact and formed a sintered neck, of radius $x_{\rm sint}$. 
The applied tensional force $F$ is zero.
(b) Geometry when a tensional force is applied. The sphere centres
have moved apart a distance $h$ and the neck begins to fail,
opening up an annular mode I crack of thickness $x_{\rm cr}$ so that
the radius of the sintered contact disc reduces to $x=x_{\rm sint}-x_{\rm cr}$.
\label{fig:fracture}}
\end{figure}

\subsection{Theory for sintering then brittle fracture}

Suppose two crystals have sintered together to form a neck of radius
$x_{\rm sint}$ with no elastic stresses present [Fig.~\ref{fig:fracture}(a)].
Eventually, this neck will break by brittle fracture (mode I loading
\cite{Zehnder}) when a tensional force $F_{\max}$ is applied.

Consider the situation when a smaller force $F < F_{\max}$
is applied, under which the neck may open up an annular
crack so that the new radius of the contact disc is $x<x_{\rm sint}$
[see Fig.~\ref{fig:fracture}(b)].
The energy per unit area of the new surface created is $\sigma_{\rm cr}$,
which for a brittle fracture will be similar to the ice/water
or ice/vacuum surface energy, but for ductile fracture will be larger,
due to plastic deformation near the crack tip. 

If we consider the scaling behavior, neglecting numerical factors
of order unity, then the energy of the new surface created is
\begin{equation}
U_{\rm surf}\sim \sigma_{\rm cr}(x_{\rm sint}^2 - x^2).
\end{equation}

Under the action of the force, the center-to-center distance increases
by $h$. The elastic deformation of the spheres is localized to a 
roughly isotropic region
of radius $x$ around the center of symmetry of the pair \cite{Hertz}.
Thus the deformed volume is of order $x^3$ and the strain of order
$h/x$. The elastic energy stored is then
\begin{equation}
U_{\rm el} \sim Y x h^2 ,
\end{equation}
where $Y$ is the elastic modulus of the crystals.

The total energy of the system (surface, elastic, and the work done
by the applied force) is thus
\begin{equation}\label{eq:Utot}
U_{\rm tot}\sim U_{\rm surf}+U_{\rm el}-Fh.
\end{equation}
For an imposed force $F$ the system will choose $h$ and $x$ to minimize 
$U_{\rm tot}$, under the constraint that $x$ cannot exceed $x_{\rm sint}$
(and by definition $x \ge 0$).

Let us define some non-dimensional parameters of the system:
\begin{equation}\label{eq:nondim_values}
\tilde{x} \equiv \frac{Yx}{\sigma_{\rm cr}}, \quad 
\tilde{h} \equiv \frac{Yh}{\sigma_{\rm cr}}, \quad
\tilde{F} \equiv \frac{YF}{\sigma_{\rm cr}^2}
\end{equation}
\begin{equation}
\Delta\tilde{U} \equiv
\frac{Y^2(U_{\rm tot}-\sigma_{\rm cr}x_{\rm sint}^{2})}{\sigma_{\rm cr}^3} \\
\end{equation}
so that from Eq.\ (\ref{eq:Utot}) the non-dimensionalized total energy is
\begin{equation}\label{eq:DeltaU}
\Delta\tilde{U} = -\tilde{x}^{2}+\tilde{h}^{2} \tilde{x} - \tilde{F}\tilde{h}.
\end{equation}

\begin{figure}
\begin{center}
\includegraphics[width=\columnwidth]{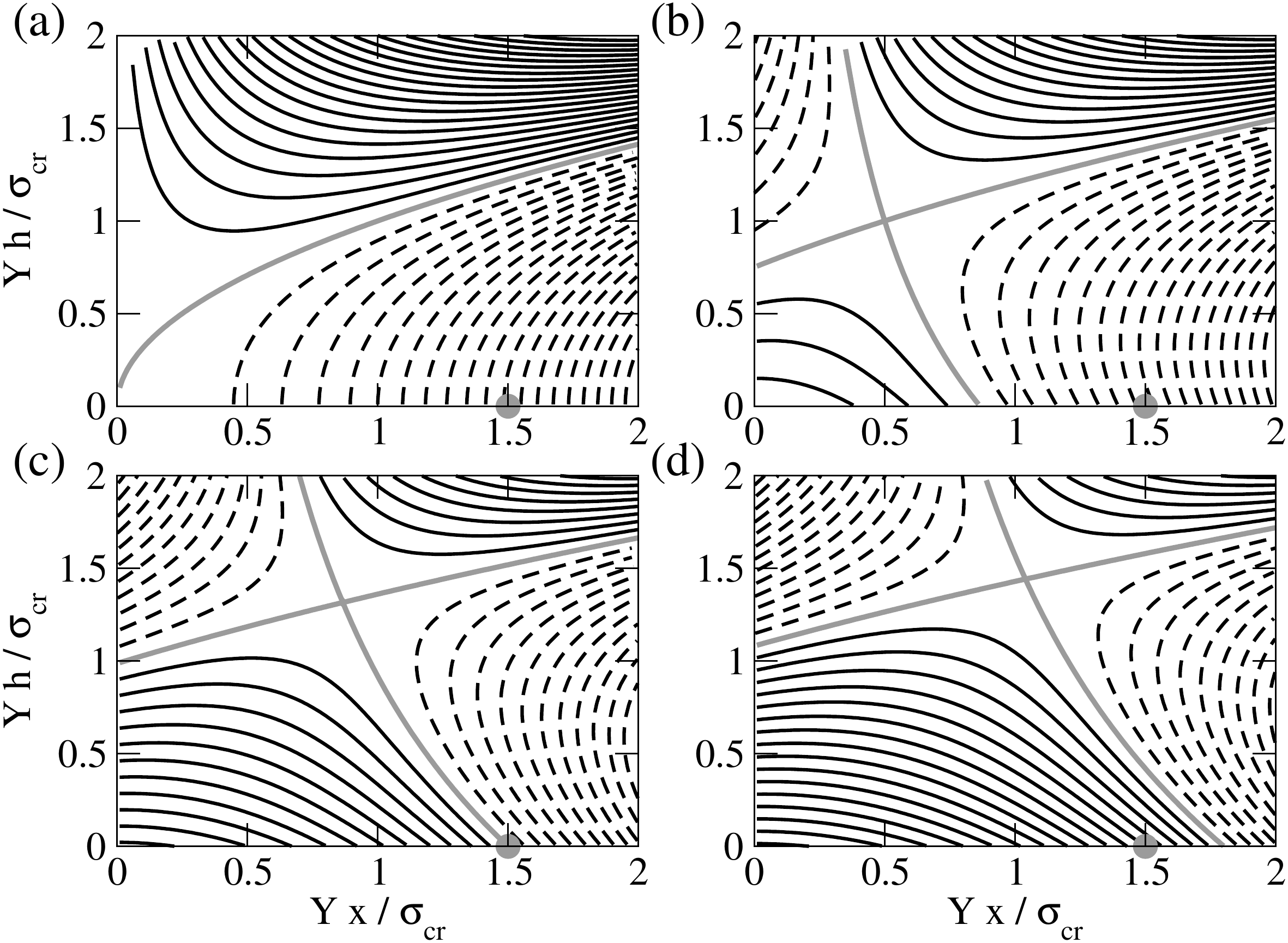}
\end{center}
\caption{Contour plots of the non-dimensionalized total energy 
$\Delta \tilde{U}$ [see Eq.\ (\ref{eq:DeltaU})] of two sintered spheres 
being pulled apart by a non-dimensionalized force $\tilde{F}$.
The initial conditions are shown by the gray circle, and correspond in
this example to $(\tilde{x},\tilde{h}) = (3/2,0)$. The contour through
the saddle point is shown in gray, and is the separatrix for two
different behaviors of the system. Solid contours have values of
$\Delta\tilde{U}$ greater than the saddle-point value, and dashed curves
have values less than this. The system will
follow a path of decreasing $\Delta\tilde{U}$, but is restricted to
$\tilde{x} \le 3/2$ at all times. (a) $\tilde{F} = 0$. The initial condition is
already the minimum energy point (under the constraint $\tilde{x}\le 3/2$).
(b) $\tilde{F}=1$. This is less than the critical value, and the minimum
energy of the system is when $\tilde{x}$ takes its maximum value of $3/2$,
and $\tilde{h}$ is greater than zero. (c) $\tilde{F}=3^{3/4}$, the critical 
value in this example. The separatrix passes through the initial conditions.
(d) $\tilde{F}=3$, greater than the critical value. There is a path,
always downhill in total energy, in which $\tilde{x} \rightarrow 0$ and
$\tilde{h} \rightarrow \infty$, so the neck between the crystals breaks
and they can separate to an arbitrarily large distance.
\label{fig:separatrix}}
\end{figure}

Consider the behavior of the system when the force $F$ is imposed.
The starting condition is $x = x_{\rm sint}$ and $h=0$. The system will then
follow a path downhill in $\Delta\tilde{U}$ in the space of 
$(\tilde{x},\tilde{h})$. This means [see Fig.~\ref{fig:separatrix}]
that provided the separatrix of
Eq.\ (\ref{eq:DeltaU}), which is the contour passing through the
saddle point, lies to the left of the starting point, the
minimum is achieved when $h$ is positive and $x = x_{\rm sint}$.
However, when $F$ increases to the point that the separatrix passes
the initial condition [Fig.~\ref{fig:separatrix}(d)], the system fails catastrophically and
$h \rightarrow \infty$ and $x \rightarrow 0$, so the neck breaks and the crystals
separate to an arbitrarily large distance.

The critical force for this to occur is $F_{\max}$, which can be found
as follows: Treating Eq.\ (\ref{eq:DeltaU}) as quantitative, the value of 
$\Delta\tilde{U}$ at the saddle point, where
$\partial\Delta\tilde{U}/\partial\tilde{x}=
\partial\Delta\tilde{U}/\partial\tilde{h}=0$ is $-(3/4)\tilde{F}^{4/3}$.
The equation for the separatrix is therefore

\begin{equation}
-(3/4)\tilde{F}^{4/3} = -\tilde{x}^{2}+\tilde{h}^{2} \tilde{x}
 - \tilde{F}\tilde{h},
\end{equation}
and this curve intersects the $\tilde{x}$-axis at
$\tilde{x}=(\sqrt{3}/2)\tilde{F}^{2/3}$. The critical force occurs
when this intersection coincides with the initial condition
$\tilde{x}=Y x_{\rm sint}/\sigma_{\rm cr}$, so restoring dimensions
using Eq.\ (\ref{eq:nondim_values})
we find (up to an unknown numerical prefactor) that
\begin{equation}\label{eq:Fmax_fracture}
F_{\max}\propto (Y\sigma_{\rm cr} x_{\rm sint}^3)^{1/2}.
\end{equation}

In order to complete this model, we need to know how the radius
$x_{\rm sint}$ of the sintered junction between two crystals
grows with contact time. Although various mechanisms of sintering are
possible \cite{German,Kingery}, including plastic flow, 
van der Waals attraction and vacancy diffusion in the crystal,
it is likely that bulk-diffusion-limited liquid-phase sintering
is the dominant process for a pair of molecularly rough \cite{Burton}
crystals brought into contact
when immersed in a solution of their melt. Even for this process,
various theories have been put forward in
the literature to describe the growth of the neck radius $x_{\rm sint}$
with contact time $t_{\rm bond}$ (where, to reiterate, in the flow, we will
choose $t_{\rm bond} = \dot{\gamma}_{\rm rms}^{-1}$).

For liquid phase sintering, consider the surface mean curvature 
$\kappa$ near the neck region of the
pair of sintering crystals. Assuming a crystal has a roughly
isotropic surface energy $\sigma_{\rm surf}$, and a latent heat
of fusion $L_f$ per unit volume, the Gibbs-Thomson effect
\cite{Meissner} states that if a flat crystal surface has an
(absolute) melting temperature
$\Theta_m$, the melting point of a curved crystal surface will
be changed by an amount
\begin{equation}\label{eq:GT}
\delta\Theta_{m} = \Theta_{m}\,\sigma_{\rm surf}\,\kappa/L_f.
\end{equation}

For our systems, the crystals are not in contact with their pure melt,
but with a solution of sucrose, so that at some temperature
$\Theta$, there will be a mass fraction $c_{\rm eq}(\Theta)$
of solute that is in equilibrium with a flat crystal surface.
This dissolution curve must be found empirically.
Assuming Eq.\ (\ref{eq:GT}) applies to the dissolution curve when
there is solute present, a simple graphical construction shows that 
at constant temperature, curvature induces a change $\delta c$ in
the equilibrium solute mass fraction in contact with ice \cite{Teardrop}:
\begin{equation}\label{eq:dc}
\delta c \approx - \Theta\, \frac{{\rm d}c_{\rm eq}}{{\rm d}\Theta}
\frac{\kappa\sigma_{\rm surf}}{L_f}.
\end{equation}
For water/sucrose systems, thermal diffusion is much faster than mass
diffusion \cite{CRC}, so the system remains isothermal.
Surface curvature induces changes in water concentration in the serum
phase, and neck growth is driven by mass diffusion down the
resulting concentration gradients.

A na{\"i}ve theory of sintering by dissolution and precipitation 
posits, on geometrical grounds, that 
\begin{equation} \label{eq:kappa}
\kappa \approx \frac{R}{x_{\rm sint}^2},
\end{equation}
and furthermore 
that the concentration difference of Eq.\ (\ref{eq:dc}) operates over a
length scale of order $R$. Therefore the prediction would be that
${\rm d}x_{\rm sint}/{\rm d}t_{\rm bond} \approx D\delta c /R$, where $D$ is 
the interdiffusivity of solute and solvent in the unfrozen serum phase. 
This can be rearranged to give
\begin{equation}\label{eq:KB_sint}
\frac{x_{\rm sint}}{R} \approx \left(
\frac{D\Theta\sigma_{\rm surf}}{R^3 L_f}\left|\frac{{\rm d}c}{{\rm d}\Theta}\right|
t_{\rm bond} \right)^{\beta} , 
\end{equation}
where $\beta=1/3$. Substituting Eq.\ (\ref{eq:KB_sint}) into 
Eq.\ (\ref{eq:Fmax_fracture}) and then Eq.\ (\ref{eq:Fmax}) leads to
the predicted exponents
\begin{eqnarray}
n_s &=& -1-\frac{3\beta}{2} , \\
n_r &=& -\frac{1}{2}-\frac{9\beta}{2} .
\end{eqnarray}
For $\beta=1/3$, this leads to predictions of $n_s = -3/2$ and $n_r = -2$.

In contrast, Courtney \cite{Courtney} argued that diffusion 
in the narrowing (wedge-shaped)
gap between the two spheres is likely to be significantly hindered by
the geometry, so that by considering diffusive trajectories and
the mass of material that gets added to the neck, it is possible to
conclude that the form of Eq.\ (\ref{eq:KB_sint}) is correct, but
with $\beta=1/5$ or $\beta=1/6$ (depending on detailed assumptions 
of the timescales involved).
Another recent theory from Farr \& Izzard \cite{Teardrop} points out 
that Eq.\ (\ref{eq:kappa}) is unlikely to be correct, 
as the narrowest region of the neck may become significantly blunted.
By considering a teardrop-shaped solitary wave solution for the 
sintering of two parallel sheets, the authors ultimately arrive at
a prediction again of the form of Eq.\ (\ref{eq:KB_sint}), but with $\beta=1/4$.

Kingery \& Berg \cite{Kingery} provide scalings for $x_{\rm sint}$
for several different sintering processes. For evaporation-condensation
they find 

\begin{equation} \label{eq:KB_1}
x_{\rm sint} \propto R^{1/3} t_{\rm bond}^{1/3},
\end{equation}
while for diffusion in the crystal

\begin{equation} \label{eq:KB_2}
x_{\rm sint} \propto R^{2/5} t_{\rm bond}^{1/5}.
\end{equation}

\begin{figure}
\begin{center}
\includegraphics[width=\columnwidth]{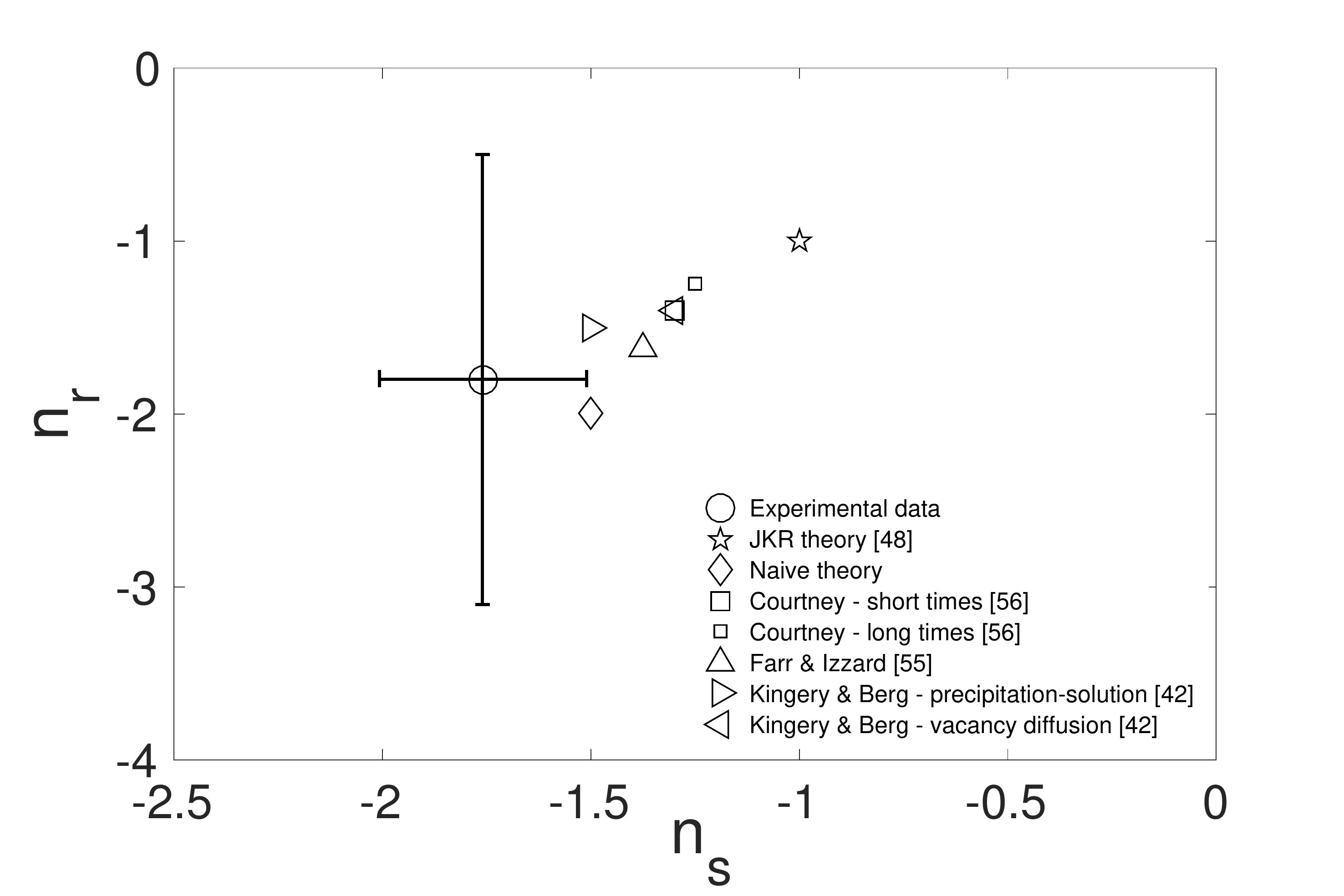}
\end{center}
\caption{Plot showing how the experimentally derived values of $n_r$ 
and $n_s$ compare to the theoretical values given in 
table~\ref{tab:predicted_exponents}.
\label{fig:exponents}}
\end{figure}

The predicted exponents for all of these mechanisms are shown in
table~\ref{tab:predicted_exponents} and are compared to the exponents 
collected from the experimental analysis (section~\ref{sec:exponent_values}) 
in Fig.~\ref{fig:exponents}. We see that there is good agreement between 
the experimental dataset collected here, and the exponents found from the 
na{\"i}ve liquid phase sintering theory and the evaporation-condensation model of 
Kingery \& Berg \cite{Kingery}. 

\begin{table*}
\begin{center}
    \begin{tabular}{lll} 
    \hline\hline
    Theoretical model\hspace{1em} & $n_s$\hspace{1em} & $n_r$ \\ \hline
    J.K.R. adhesive spheres \cite{JKR} & $-1$ & $-1$\\
    Na{\"i}ve theory of fracture \& liquid phase sintering	& $-3/2$ & $-2$ \\    
    Fracture \& liquid phase sintering by diffusion, short times (Courtney \cite{Courtney})       & $-13/10$ & $-7/5$\\
    Fracture \& liquid phase sintering by diffusion, long times (Courtney \cite{Courtney})       & $-5/4$   & $-5/4$\\
    Fracture \& liquid phase sintering (Farr \& Izzard \cite{Teardrop}) & $-11/8$  & $-13/8$\\ 
    Fracture \& sintering by evaporation-condensation (Kingery \& Berg \cite{Kingery}) & $-3/2$   & $-3/2$\\
    Fracture \& sintering by vacancy diffusion (Kingery \& Berg \cite{Kingery})	& $-13/10$ & $-7/5$\\ \hline
    Experimental results & $-1.76\pm0.25$ & $-1.80\pm1.30$    \\
    \hline\hline
    \end{tabular}
    \caption{\label{tab:predicted_exponents}Theoretical predictions for the
exponents $n_s$ and $n_r$ in Eq.\ (\ref{eq:eta_power_law}) from different
literature sources, as analyzed in section~\ref{sec:theory}, compared to the
experimental exponents from this study.}
\end{center}
\end{table*}

\section{Conclusions and Implications}

We have found that the viscosity of ice-sucrose mushes has a power-law dependence on shear rate and crystal size. These powers can be explained due to the formation of clusters of crystals by liquid phase sintering and break-up of these clusters by brittle fracture under flow. We have observed under long timescales the crystals ripen under shear, with a power of time that is smaller than any published theory. We interpret this as possibly due to abrasion or effects of cluster break-up on the crystals.


We believe that the present work, on a model system with sub-spherical,
sintering crystals, represents a step towards understanding the rheology
of a much broader class of materials which are of great technological and 
social importance. This class of material includes volcanic and cryogenic lavas, 
frazil ice, frozen foods and salt slurries.

The phenomenology of ice-sucrose systems should be directly relevant to 
the behavior of cryogenic lavas, which have been discovered widely throughout
the moons and minor planets beyond Mars; for example Ceres \cite{Ceres},
Ganymede \cite{Ganymede}, Titan, Enceladus \cite{Zhong}, Triton \cite{Triton} and
Pluto \cite{Pluto}.
These lavas have various ices present as crystal phases, therefore one 
would expect liquid phase sintering as suggested here will be relevant 
to the features produced by such flows \cite{Lopes}.
These results will be of more limited importance to the study of terrestrial 
lavas, as in general the rheology of such flows can be described by a 
modified Krieger-Dougherty relationship \cite{Vona}. 

For large bodies of magma which contain reservoirs of crystal mush, the 
situation is different. Liquid phase sintering has been predicted to occur 
in granitic mushes with a high degree of partial melt and confirmed through 
dihedral angles between quartz-feldspar and quartz-quartz grain junctions 
\cite{Jurewicz}. There is also extensive evidence of solution-precipitation
sintering dynamics in olivine-basalt aggregates \cite{CooperKohlstedt1,CooperKohlstedt2}.

Knowing viscosity is of crucial importance in understanding 
the behavior of magmatic hazards. Crystal-rich ignimbrites (from highly 
voluminous, explosive eruptions) are observed in the rock record \cite{Huber}, 
with the crystals expected to come from a long-lived stored mush reservoir 
\cite{Cooper} which has been triggered due to a change in the thermal state 
of the reservoir \cite{Huber}. Although the Krieger-Dougherty relationship is 
often used in the literature to describe eruptible magmas 
with $\phi \approx 0.5-0.6$ \cite{Huber}, the yield stress behavior just 
noted and the shear-rate and time dependence that we have uncovered in the 
present work suggest that eruptibility of these magmas is being overstated 
(and the viscosity vastly underestimated) by the Krieger-Dougherty relation.

Moving beyond subspherical crystals, frazil ice forms due to turbulent mixing of supercooled salt water, often in polynyas near ice shelves \cite{Martin}. The ice crystals formed are needle shaped, and often stick together in clots. We believe the rheology of mushes of high aspect ratio crystals is both of critical importance (since it applies to many magmatic systems as well as frazil ice) but has received little study in the literature. Such systems bring added complexity and experimental challenges from the possibility for flow induced alignment. 


\acknowledgments

The authors gratefully acknowledge William Frith, Jerome Neufeld and
Marian Holness for many useful discussions. AJG acknowledges a NERC CASE studentship
award [grant number D07591442: MA-2013-00657].

\end{document}